\documentclass[nopubldata]{lpor2014}

\usepackage{siunitx}
\graphicspath{{figs/}}
\usepackage{color}
\usepackage{overpic}
 
\usepackage[overlay,absolute]{textpos}
\newcommand\PlaceText[3]{%
	\begin{textblock*}{10in}(#1,#2)
		#3
	\end{textblock*}
}%

\begin{document}
	\titlefigure{abstract_graphic}
	
	\abstract{
With growing interest in the mid-infrared spectral region, dysprosium has recently been revisited for efficient high-performance infrared source development.
Despite historically receiving less attention than other rare earth ions, in recent years lasers utilising the dysprosium ion as the laser material have set record mid-infrared performance, including tunability from 2.8 to \SI{3.4}{\micro\m} (in addition to \SI{4.3}{\micro\m} lasing), CW powers exceeding 10~W, greater than 73\% slope efficiencies and even ultrafast pulsed operation.
In this review, we examine the unique energy level structure and spectroscopy of the dysprosium ion and survey the major developments that have led to this resurgence of interest and subsequent record mid-infrared laser performance.
We also highlight mid-infrared applications of emerging dysprosium lasers, in addition to surveying the many opportunities that lie ahead.}
	
	\title{Dysprosium mid-infrared lasers: Current status and future prospects}
	
	\titlerunning{}
	\author{M. R. Majewski\inst{*}, R. I. Woodward, and S. D. Jackson}
	\authorrunning{}
	\institute{%
		MQ Photonics, School of Engineering, Macquarie University, North Ryde, NSW 2109, Australia
	}
	\mail{\email{matthew.majewski@mq.edu.au}}

	\maketitle
	
	\PlaceText{40mm}{9mm}{Laser \& Photonics Reviews \textbf{14}, 1900195 (2020); https://doi.org/10.1002/lpor.201900195}

	\section{Introduction}
	\label{sec:intro}

The mid-infrared (MIR) wavelength range, spanning the \SI{2.5}{\micro\m} to \SI{30}{\micro\m} region of the electromagnetic spectrum, is a current scientific and technological frontier, which holds great potential for new photonic applications.
This is due to the MIR region containing characteristic rotational/vibration frequencies of molecular bonds, which leads to resonant absorption of light that makes the MIR vital to technologies in medicine, sensing, and materials processing.

To date, research activity into MIR rare-earth source development has primarily focussed on the trivalent erbium (Er$^{3+}$) and holmium (Ho$^{3+}$) ions.
While promising results have been demonstrated, particularly at \SI{3}{\micro\m}, this large body of work has overlooked another highly promising ion: trivalent dysprosium (Dy$^{3+}$).
Dysprosium is relatively unique among rare earth ions, notably offering emission around \SI{3}{\micro\m} from the lowest energy transition (thus facilitating simple in-band pumping for high efficiency), with significantly wider emission cross sections than alternative ions.
It also offers a number of transitions for even longer wavelength emission.
While dysprosium was first used as an active laser material in 1973, 13 years after the pioneering demonstration of the first laser, the ion received little subsequent research attention compared to other rare earths: only a handful of Dy$^{3+}$-doped lasers have been demonstrated and its potential remains largely untapped.

Fortunately, following a number of developments, this situation is now changing and in recent years dysprosium has experienced a resurgence of interest: the number of reported dysprosium lasers has quadrupled in the last decade and performance is now rivalling, and even surpassing, alternative MIR sources.
It is therefore timely to review the current understanding of the dysprosium ion and evaluate the state of the art in Dy$^{3+}$-doped laser technology.
Section~\ref{sec:spectroscopy} starts with a detailed description of dysprosium spectroscopy, summarising measured spectroscopic quantities in various crystal and glass hosts, alongside co-doping opportunities.
Section~\ref{sec:lasers} then considers Dy$^{3+}$-doped lasers and novel cavity design, including tabulation of all reported MIR lasers to date, to the best of our knowledge.
Finally, we conclude with a critical evaluation of dysprosium as a MIR gain medium and comment on future opportunities in Section~\ref{sec:outlook}.

	\section{Spectroscopy}
	\label{sec:spectroscopy}

	\subsection{Energy level structure}
\begin{figure}
	\includegraphics*{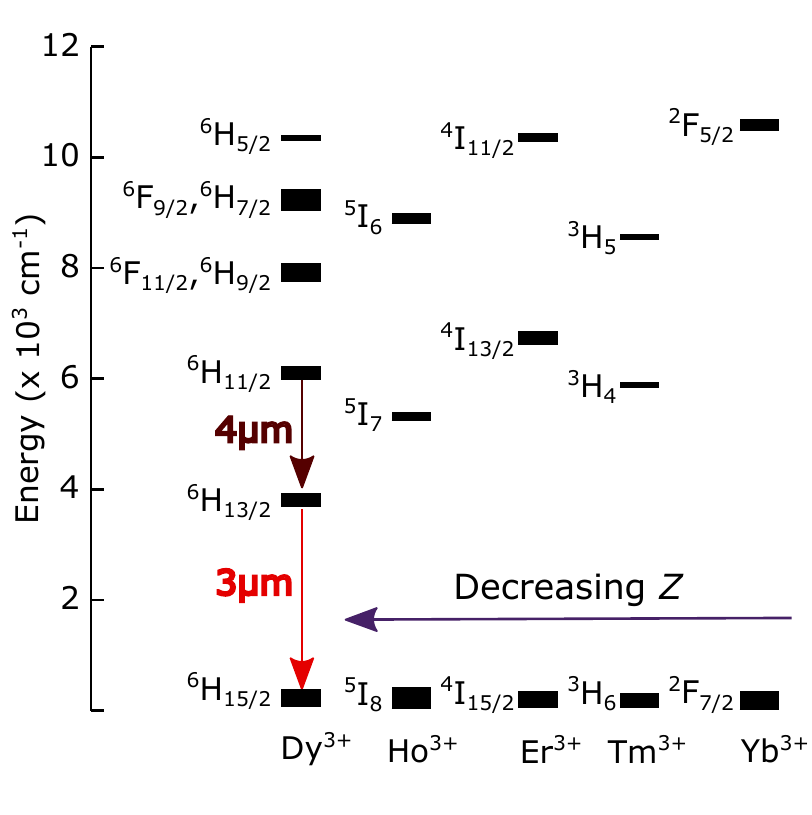}
	\caption{Low-lying energy level manifolds of several trivalent rare-earth elements in LaCl$_3$ (adapted from Ref.~\cite{Dieke1961}). From right to left is decreasing atomic number $Z$ showing a trend towards reduction in energy of the first few excited states. Possible emission from dysprosium is indicated.}
	\label{fig-RE-levels}
\end{figure}

The trivalent dysprosium ion is relatively unique in the set of rare-earth dopants.
With the significant exception of Nd$^{3+}$, the most heavily researched and commercially successful optically active ions are from the high atomic number end of the lanthanide series.
Specifically, Yb$^{3+}$ fiber lasers emitting around \SI{1}{\micro\m} are the modern standard for high power industrial lasers, while Er$^{3+}$-doped fiber amplifiers at \SI{1.55}{\micro\m} underpin vast telecommunication networks.
Both of these lasers operate on transitions between the Stark level manifolds of the first excited and ground states. 
As the simplified energy level diagram in Figure~\ref{fig-RE-levels} shows, the trend towards a lower energy (longer emission wavelength) first excited state continues with decreasing atomic number ($Z$): with Tm$^{3+}$ and Ho$^{3+}$ both capable of emission to the ground state centered around 1.9 and \SI{2.1}{\micro\m}, respectively. 

Dysprosium exhibits several low-lying energy levels, and according to the noted trend with successively smaller $Z$ is the first element with a first excited state $^6H_{13/2}$ corresponding to emission in the MIR spectral region, at nominally \SI{3}{\micro\m}.
Additionally, the second excited state $^6H_{11/2}$ lies nominally only 6000~\si{\per\cm} above the ground state, with resultant emission to the first excited state at \SI{4.3}{\micro\m}.

Further decrease in $Z$ of rare-earth ions largely maintains the trend towards multiple low-energy excited states (with the exception of Gd$^{3+}$), though emission from these levels is practically quenched completely by multiphonon absorption in a vast majority of hosts.
Thus dysprosium-doped lasers offer the distinctive potential for sources of high quantum efficiency MIR light.

The highest energy level of the group of Dy$^{3+}$'s states below 14000~\si{\per\cm}, $^6F_{3/2}$, is separated from the next excited state, $^4F_{9/2}$, by an energy gap of $\sim$8000~\si{\per\cm}. 
This $^4F_{9/2}$ level is metastable and largely immune from the influence of multiphonon relaxation, leading to some interest in dysprosium laser sources emitting in the visible spectral range~\cite{Kaminskii2002,Parisi2005,Macalik1998,Gruber1998}.
Specifically, the relatively recent commercial availability of GaN blue laser diodes has enabled demonstration of dysprosium lasers operating on the ${^4F_{9/2}\rightarrow^6H_{13/2}}$ transition around 580~nm~\cite{Limpert2000,Fujimoto2010,Bowman2012,Bolognesi2014}.

The multiple low-energy dysprosium states also historically generated some interest as a saturable absorber candidate to be used in Q-switched laser systems~\cite{Sardar2004,Seltzer1996}, though this has been superseded by successful alternatives, for example semiconductor saturable absorber mirrors (SESAMs). 
Further optical property study of the dysprosium ion was motivated by emission from the $^6H_{9/2} - ^6F_{11/2}$ multiplet at \SI{1.3}{\micro\m}.
This ground-state-terminated transition showed some promise as an amplifier for telecommunications systems in the O-band~\cite{Page1997,Tkachuk1999a,Tanabe1995,Ballato1997,Amplifiers1998}, though this has also been superseded, in this case by large-scale transition to the C-band around \SI{1.55}{\micro\m}, driven by the emergence of EDFAs.
\begin{figure*}
	\begin{overpic}{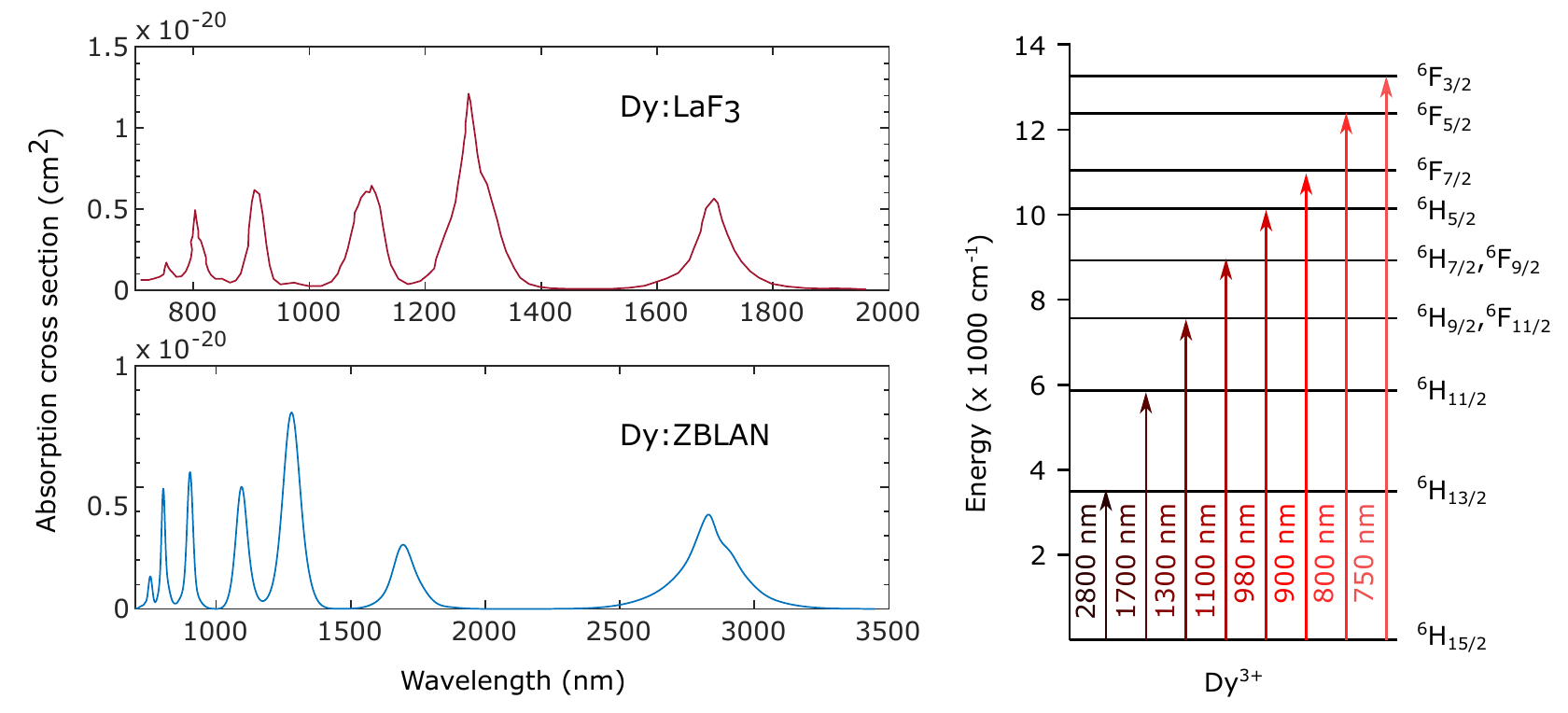}
		\put(10,40){(a)}
		\put(10,20){(b)}
		\put(60,40){(c)}
		\end{overpic}
	\caption{Measured Dy$^{3+}$ absorption cross sections in representative a) crystalline (LaF$_3$)~\cite{Li2017}, and b) glass (ZBLAN)~\cite{Piramidowicz2008b} hosts. c) Energy level diagram indicating relevant ground state absorption transitions. Note that 980~nm ground state absorption to the $^6H_{5/2}$ level is included for completeness, but is sufficiently weak that it is not readily observed experimentally.  }
	\label{fig-abs}
\end{figure*}
Additionally, the ${^6F_{11/2},^6H_{9/2}\rightarrow^6H_{11/2}}$ transition, producing emission around \SI{5.5}{\micro\m} has been experimentally investigated by a limited number of authors~\cite{Shestakova2007,Jelinkova2016}.
This transition is strongly self-terminating, which prohibits steady-state continuous wave (CW) lasing; only short pulse emission is possible with notably low efficiency of $<1$\%.
The luminescence efficiency of emission from the upper state, and the self-terminating nature of the transition are both negatively impacted by the strong influence of multiphonon decay; a dominant process even in low-phonon energy crystals.

Aside from these few alternative aims, the majority of interest in dysprosium-doped media has been in the context of MIR light generation, and the effort to exploit the aforementioned potential for high quantum efficiency laser sources either at 3 or \SI{4}{\micro\m}.
The multiple ground state absorption transitions in the near infrared (see Figure~\ref{fig-abs}) offer a variety of routes to accumulating excited state population.
Specifically, dysprosium exhibits absorption features centered around \numlist{0.8;0.9;1.1;1.3;1.7;2.9}~\si{\micro\m}.
With the exception of \SI{2.9}{\micro\m}, which populates $^6H_{13/2}$ directly, each of these pump wavelengths may be considered to only populate the $^6H_{11/2}$ and $^6H_{13/2}$ levels as transitions between each of the states above $^6H_{11/2}$ are strongly quenched by non-radiative decay in virtually all hosts of practical significance.
In a large subset of considered hosts, $^6H_{11/2}$ and $^6H_{13/2}$ are also strongly coupled by multiphonon relaxation, rendering all of the noted pump absorption wavelengths viable candidates for producing emission around \SI{3}{\micro\m}.

The primary driver in suitable host selection for MIR laser sources is minimizing non-radiative decay, i.e. choosing low phonon energy materials. 
This requirement is readily satisfied in several crystalline hosts, suitable for bulk laser geometries, although glass materials must also be considered that can be drawn into high-quality fiber for fiber laser systems.
Thus we separate both the subsequent spectroscopic discussion and the review of laser demonstration by host designation, either crystalline or glass.

	\subsection{Dy$^{3+}$-doped crystals}
\begin{table}
	\centering
	\caption{Experimentally observed and calculated Stark level positions (\si{\per\cm}) of Dy$^{3+}$ ground and first excited states in various crystalline hosts.}
	\label{Stark_table}
	\begin{tabular}{@{}ccccc@{}}
		\toprule
		Level     & BaY$_2$F$_8$ & LiYF$_4$ & LaF$_3$ (calc) & YAG (calc) \\  \midrule
		$^6H_{15/2}$ &      0       &    0    &       22       &     0      \\
		&     7.5      &   14    &       26       &     57     \\
		&     39.5     &   41    &       79       &    101     \\
		&      48      &   62    &      106       &    172     \\
		&     67.5     &   70   &      195       &    235     \\
		&     109      &   110   &      216       &    474     \\
		&     216      &   344   &      323       &    517     \\
		&     592      &   375   &      335       &    732     \\
		&              &         &                &            \\
		$^6H_{13/2}$ &     3507     &  3493   &      3493      &    3563    \\
		&     3535     &  3516   &      3579      &    3595    \\
		&     3559     &  3534   &      3616      &    3695    \\
		&     3581     &  3579   &      3636      &    3717    \\
		&     3638     &  3632   &      3637      &    3780    \\
		&     3683     &  3717   &      3671      &    3819    \\
		&     3836     &  3726   &      3679      &    3946    \\
		&              &         &                &            \\
		$^6H_{11/2}$ &     5819     &  5881   &      5873      &    5932    \\
		&     5861     &  5875   &      5910      &    5952    \\
		&     5901     &  5925   &      5913      &    6042    \\
		&     5957     &  5935   &      5931      &    6084    \\
		&     6022     &  5977   &      5968      &    6100    \\
		&     6052     &  5987   &      6021      &    6106    \\
		&              &         &                &            \\
		Reference   &      \cite{Johnson1973b}        &   \cite{Davydova1977a}      &   \cite{Carnall}             &	\cite{Grunberg1969}		\\    \bottomrule
	\end{tabular} 
\end{table}
Crystalline hosts for laser active ions are often desirable due to their high thermal conductivity and optical quality, as well as the generally large transition cross sections.
One of the most common crystals used in solid state laser development is yttrium aluminium garnet, YAG.
However, as the maximum phonon energy of YAG is approximately 850~\si{\per\cm}, the multiphonon absorption edge is between 4 and \SI{5}{\micro\m}, rendering it not the most desirable host for MIR laser operation.
To enable dysprosium laser action on both the ${^6H_{13/2}\rightarrow^6H_{15/2}}$ and ${^6H_{11/2}\rightarrow^6H_{13/2}}$ transitions, alternative crystals are generally necessary. Focus historically has been on fluoride crystals such as BaY$_2$F$_8$ and LaF$_3$.

Due to the crystalline nature of the host and the minimal impact of inhomogeneous broadening, rare-earth doped crystals typically exhibit well defined transitions, often allowing for direct experimental determination of Stark level positions.
Analytical assignment of Stark level energies is also possible, and though dysprosium has attracted proportionally less research attention than other rare earths, it was included in some of the original comprehensive studies concerning rare-earth-doped LaF$_3$~\cite{Dieke1961,Carnall1968,Carnall,Neogy1988,wortman1976rare}.
Numerical results from a representative calculation are presented in Table.~\ref{Stark_table}.
Experimentally determined Stark level positions in both BaY$_2$F$_8$ and LiYF$_4$ at cryogenic temperatures under ultraviolet excitation~\cite{Davydova1977a,Johnson1973b} are also presented.
While there is some variation in splitting with the differing crystal fields of the various hosts, all of the fluorides show a generally reduced amount of splitting as compared to the comparably stronger crystal field of an oxide host such as YAG.

Transition strengths typically are determined through a Judd-Ofelt analysis, generating three phenomenological intensity parameters $\Omega_{\lambda}$ from measured absorption data.
Values obtained for these parameters in some of the crystalline hosts used for laser emission are shown in Table~\ref{JO_table_crystal}. 
The computed transition strengths can be used to determine radiative lifetimes and branching ratios between manifolds, although since they are integrated values, they cannot directly give absorption  / emission cross section values without independently measuring the spectral shape~\cite{Hehlen2013,Walsh2006}.
\begin{table}[!htbp]
	\centering
	\caption{Judd-Ofelt intensity parameters ($\Omega_{\lambda}$ , \SI{e-20}{\cm^2}) in various crystal materials }
	\label{JO_table_crystal}
	\begin{tabular}{@{}ccccc@{}}
		\toprule
		Host            & $\Omega_2$ & $\Omega_4$ & $\Omega_6$ &         Reference          \\ \midrule
		LaF$_3$      &    1.1     &    1.4     &    0.9     &       \cite{Xu1984}        \\
		BaY$_2$F$_8$    &    1.52    &    2.33    &    3.67    &     \cite{Parisi2005}     \\
		LiYF$_4$~(YLF)  &    1.6     &    1.87    &    2.52    &      \cite{Brik2004}       \\
		\bottomrule
	\end{tabular}
\end{table} 

Room temperature measured absorption data for dysprosium in the representative host LaF$_3$ is presented in Figure~\ref{fig-abs}(a). 
While there remains some structure indicative of transitions between individual Stark levels, generally the absorption features present as broad and nearly smooth.
There are mostly minor differences in the absorption characteristics amongst  various crystal host which have been investigated.
The only notable exception is YLF, where the dominant peak is measured to be the 800~nm absorption to the $^6F_{5/2}$ level, with a nominal magnitude of \SI{1.4e-20}{\cm^2}~\cite{Brik2004,Ivanova1999}, comparable to the \SI{1.3}{\micro\m} cross section in LaF$_3$~\cite{Li2017}.
While not present in the dataset shown in Figure~\ref{fig-abs}(a), the $^6H_{13/2}$ absorption around \SI{3}{\micro\m} is nominally of similar magnitude to $^6H_{11/2}$, i.e. \SI{6e-21}{\cm^2}. 

\begin{figure}[!htbp]
	\begin{overpic}{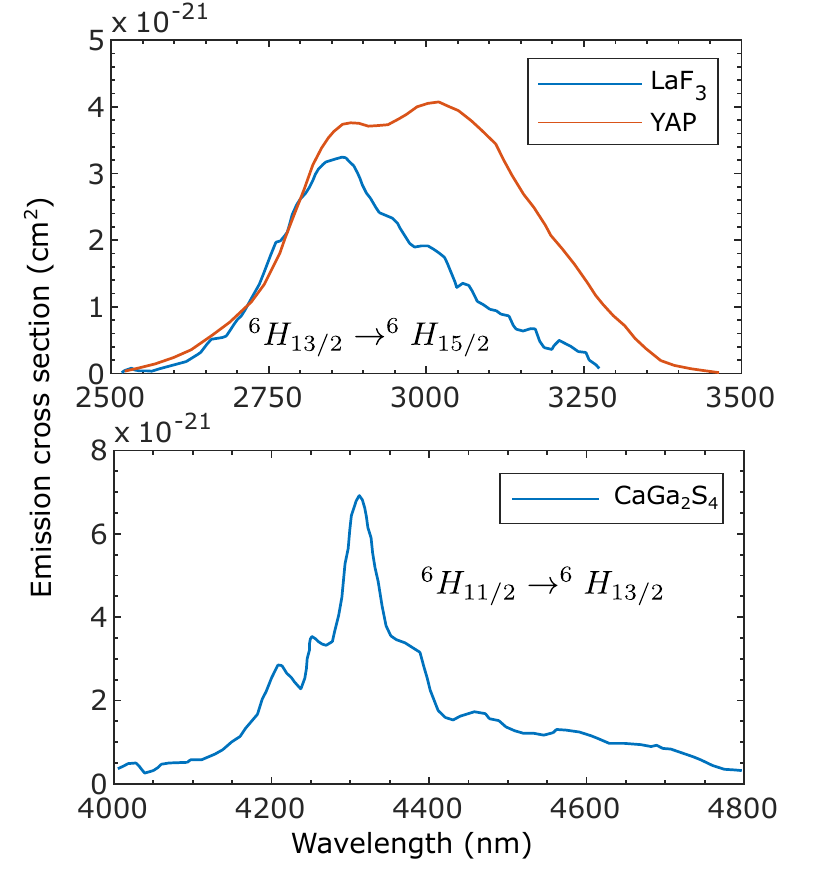}
		\put(15,90){(a)}
		\put(15,42){(b)}
		\end{overpic}
	\caption{Measured emission cross sections of the a) ${^6H_{13/2}\rightarrow^6H_{15/2}}$ transition in LaF$_3$ and YAlO$_3$ (YAP)~\cite{Li2017,Wang2014}, and the b)  ${^6H_{11/2}\rightarrow^6H_{13/2}}$ transition in a sulfide crystal (CaGa$_2$S$_4$)~\cite{Nostrand1999}. }
	\label{fig-crystal-emission}
\end{figure}

Measured emission cross section spectra from the ${^6H_{13/2}\rightarrow^6H_{15/2}}$ transition are seen in Figure~\ref{fig-crystal-emission}(a) for both LaF$_3$ and YAP (YAlO$_3$).
Emission broadly spans from \SI{2.65}{\micro\m} to at least \SI{3.2}{\micro\m} in LaF$_3$, notably covering the \SI{2.9}{\micro\m} water absorption peak.
Due to the increased ground state splitting in the oxide host (YAP), emission extends somewhat further to \SI{3.4}{\micro\m}.
Significantly, this transition is capable of emission well beyond Er and Ho alternative \SI{3}{\micro\m} laser candidates.
The peak cross section magnitude differs between the two hosts by 20\%, while data for additional hosts in Table~\ref{emission_table_crystal} indicate that larger values are possible.
Greater than 50\% increase over YAP is found in a chloride-based host, while very recently yet further improvement was determined for cubic sesquioxide Lu$_2$O$_3$, though this material is also shown to suffer from substantial increase in the detrimental effect of multiphonon relaxation~\cite{Heuer2018}. 

Comparatively less spectroscopic study has focused on the ${^6H_{11/2}\rightarrow^6H_{13/2}}$ transition around \SI{4.3}{\micro\m}.
Available data is confined primarily to sulfide and chloride hosts, which as seen in Table~\ref{emission_table_crystal} exhibit essentially equivalent values for peak emission cross section.
The spectral shape of emission as measured in CaGa$_2$S$_4$ is presented in Figure~\ref{fig-crystal-emission}(b),  with emission spanning nearly 800~nm and a peak centered nominally at \SI{4.3}{\micro\m}.
There is strong overlap of this transition with atmospheric absorption by carbon dioxide (CO$_2$).
It should be noted that as cross sections are generally calculated from measured integrated emission spectra, the contribution of CO$_2$ absorption directly affects accurate calculation, thus values obtained could be subject to a fair degree of uncertainty.

\begin{table}[!htbp]
	\centering
	\caption{Peak emission cross sections ($\sigma_e$ , \SI{e-21}{\cm^2}) of the low-lying MIR Dy$^{3+}$ transitions in various crystals. Transitions terminate on the the ground ($^6H_{15/2}$) and first excited state ($^6H_{13/2}$), respectively. }
	\label{emission_table_crystal}
\begin{tabular}{@{}cccc@{}}
	\toprule
	    Host      & $^6H_{11/2}\rightarrow^6H_{13/2}$ & $^6H_{13/2}\rightarrow^6H_{15/2}$ &      Reference       \\ \midrule
	CaGa$_2$S$_4$ &                 7                 &                                   & \cite{Nostrand1999}  \\
	KPb$_2$Cl$_5$ &                6.9                &                6.1                & \cite{Nostrand2001}  \\
	     YAP      &                                   &                 4                 &   \cite{Wang2014}    \\
	   LaF$_3$    &                                   &                3.2                &    \cite{Li2017}     \\
	 Lu$_2$O$_3$  &                                   &                6.7                 &   \cite{Heuer2018}   \\
	        \bottomrule
\end{tabular}
\end{table} 

Judd-Ofelt analysis further allows determination of excited state radiative lifetimes. 
In practice, fluorescence lifetimes are often observed to be smaller than the calculated value due to the influence of multiphonon relaxation (MPR).
Calculated and direct experimentally measured values for lifetimes of both the $^6H_{11/2}$ and $^6H_{13/2}$ state are presented for various crystals in Table~\ref{lifetime_table_crystal}.
While strong variation in calculated (i.e. radiative) lifetime is observed between different hosts, it is noted that branching ratios are approximately host-independent, e.g. the ${^6H_{11/2}\rightarrow^6H_{13/2}}$ branching ratio is typically $\sim$6\% (thus, $\sim$94\%~~${^6H_{11/2}\rightarrow^6H_{15/2}}$ branching ratio).

Due to a lack of Dy$^{3+}$ spectroscopic data in the literature, it is not always possible to fully compare the behavior of different crystal hosts.
However, one constant is that the $^6H_{13/2}$ lifetime is always longer than that of $^6H_{11/2}$, rendering the transition ``self-terminating''.
Pulsed operation remains possible, with the upper bound of useful pumping duration being approximately proportional to the difference between upper and lower state lifetimes.
As both lifetimes are of similar order of magnitude across hosts, pump pulses of duration less than nominally 1~ms can potentially produce population inversion.

Where direct measurements exist, experimentally observed values of fluorescence lifetime are generally seen to be substantially smaller than the calculated radiative lifetime.
This is to be expected because measurements in actuality reflect the combination of radiative lifetime and contribution of  multiphonon non-radiative relaxation, which is not included in Judd-Ofelt analysis.
Multiphonon relaxation is a function both of the separation in energy between two states, and the phonon energy of the host.
For example, the \SI{613}{\per\cm} maximum phonon energy of Lu$_2$O$_3$ is responsible for the three orders of magnitude reduction in the observed lifetime from the value calculated by Judd-Ofelt analysis~\cite{Heuer2018}.
Lanthanum fluoride (LaF$_3$) possesses a considerably lower phonon energy of \SI{350}{\per\cm}, with a correspondingly reduced influence of multiphonon decay on the observed $^6H_{13/2}$ lifetime.
Laser oscillation threshold scales inversely with upper state lifetime, imposing a stringent requirement on host phonon energy, particularly for \SI{4.3}{\micro\m} emission.

\begin{figure}[!htbp]
	\includegraphics{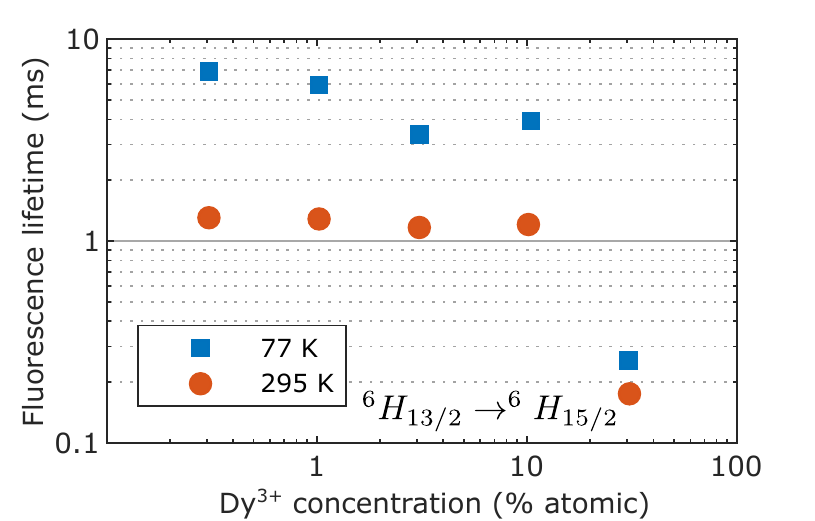}
	\caption{Measured \SI{3}{\micro\m} fluorescence lifetime as a function of Dy$^{3+}$ concentration in a BaY$_2$F$_8$ host at cryogenic and room temperatures. Strong concentration quenching is observed above 10\%. Data from Ref.~\cite{Johnson1973b}. }
	\label{fig-quenching-crystal}
\end{figure}

Multiphonon effects scale directly with host temperature, and can be substantially reduced under cryogenic cooling.
Fluorescence lifetime measurements of the ${^6H_{13/2}\rightarrow^6H_{15/2}}$  transition in BaY$_2$F$_8$ at room temperature and with liquid nitrogen cooling (\SI{77}{\kelvin}) are presented in Figure~\ref{fig-quenching-crystal}.
Considerable reduction in lifetime is observed at room temperature due to the increased thermal phonon population.
This data additionally demonstrates a pronounced concentration induced quenching of the lifetime, a phenomena which is seldom considered in the literature.
Analysis of the energy level structure indicates that there are no ion-ion interactions appreciably resonant enough to induce quenching of the $^6H_{13/2}$ level, thus the behaviour is attributed to the presence of an impurity in the host~\cite{Nostrand2001,Johnson1973b}.
Similar findings in glass hosts, as will be discussed, identify the impurity as OH$^{-}$.
Concentration quenching and general energy transfer effects on the $^6H_{11/2}$ level have not yet been observed or considered analytically in particular detail in the existing literature.
\begin{table}[!htbp]
	\centering
	\caption{Calculated radiative lifetimes ($\tau_{calc}$) and experimentally observed fluorescence lifetimes ($\tau_{meas}$) (ms) of Dy$^{3+}$ in selected crystalline hosts.}
	\label{lifetime_table_crystal}
\begin{tabular}{@{}cccccc@{}}
	\toprule
	               & \multicolumn{2}{c}{$^6H_{11/2}$} & \multicolumn{2}{c}{$^6H_{13/2}$} &                       \\
	     Host      & $\tau_{calc}$ &  $\tau_{meas}$   & $\tau_{calc}$ &  $\tau_{meas}$   &       Reference       \\ \midrule
	BaYb$_2$F$_8$  &        0.2       &               &           1.9    &       0.9        &    \cite{Mak1982}     \\
	CaGa$_2$S$_4$  &         4.16      &      2.9       &     13.2          &     6.9         &  \cite{Contact2012}  \\
	RbPd$_2$Cl$_5$ &               &       5.6        &               &        13        & \cite{Shestakova2007} \\
	     YLF       &     14.8      &                  &     62.5      &                  &  \cite{Ivanova1999}   \\
	     YSGG      &     19.7      &                  &      66       &                  &   \cite{Sardar2004}   \\
	     YAP       &               &                  &     15.3      &       8.9        &    \cite{Wang2014}    \\
 KPb$_2$Br$_5$& & &6.9 &3 &\cite{Hommerich2006} \\
	   LaF$_3$     &               &                  &      22       &       9.7        &     \cite{Li2017}     \\
	 Lu$_2$O$_3$   &               &                  &      28       &       0.05       &   \cite{Heuer2018}    \\ 
	  
	 \bottomrule
\end{tabular} 
\end{table}	

	\subsection{Dy$^{3+}$-doped glasses}
While crystalline hosts possess several inherent properties which are advantageous for laser operation, a major disadvantage is that they cannot be drawn into high-quality low-loss optical fibers, restricting their use to bulk laser geometries. 
Strong demand exists for the development of MIR fiber lasers, however, thus driving considerable research effort into dysprosium-doped glass materials which are suitable for fiber fabrication.

\begin{table}[!htbp]
	\centering
	\caption{Judd-Ofelt intensity parameters ($\Omega_{\lambda}$ , \SI{e-20}{\cm^2}) in selected glass host materials. }
	\label{JO_table_glass}
	\begin{tabular}{@{}ccccc@{}}
		\toprule
		Host            & $\Omega_2$ & $\Omega_4$ & $\Omega_6$ &         Reference          \\ \midrule
	
		ZBLAN            &    1.86    &    1.42    &    2.73    &    \cite{Wetenkamp1992}    \\
		ZBLAN            &    3.03    &    1.32    &    2.06    & \cite{mcdougall1994judd-a} \\
		ZBLAN            &    3.12    &    0.95    &    2.3     &  \cite{Piramidowicz2008b}   \\
		ZBLA            &    2.65    &    1.29    &    1.72    &  \cite{mcdougall1994judd}  \\
		ZBLA            &    3.22    &    1.35    &    2.38    &      \cite{Adam1988}       \\
		ZBLALi           &    2.7     &    1.8     &     2      &  \cite{orera1988optical}   \\
		HfF$_4$           &    3.12    &    2.07    &    2.48    &      \cite{Cases1991}      \\
		TeO$_2$           &    4.28    &    1.32    &    2.53    &    \cite{Hormaldy1979}     \\
		Ge$_{30}$As$_{10}$S$_{60}$ &   10.53    &    3.17    &    1.17    &       \cite{Heo1996}       \\
		Ge$_{25}$Ga$_{5}$S$_{70}$  &    11.9    &    3.58    &    2.17    &       \cite{Wei1994}       \\ \bottomrule
	\end{tabular}
\end{table}

In terms of low propagation loss in passive fiber, by far the most successful to date has been heavy metal fluoride fiber, specifically ZBLAN (ZrF$_4$-BaF$_2$-LaF$_3$-AlF$_3$-NaF), which is available commercially with optical loss of $<$\SI{10}{\decibel\per\kilo\meter}.
Indeed, nearly the entirety of Dy$^{3+}$-doped glass laser demonstrations have been ZBLAN fiber systems.
However, several alternative doped glasses have been investigated, most notably those of the chalcogenide family, such as the sulfides and selenides.
These glasses have substantially larger refractive indices, offering the promise of increased transition cross sections, but it is their reduced phonon energy as compared to ZBLAN which is of paramount importance in the scope of dysprosium fiber lasers.
Transparency in ZBLAN is practically limited to wavelengths below \SI{4}{\micro\m} due to exponentially increasing multiphonon absorption, with a nominal maximum phonon energy of \SI{600}{\per\cm}.
Whereas chalcogenide glasses exhibit maximum phonon energies as low as \SI{300}{\per\cm}~\cite{Sojka2017}.

Judd-Ofelt intensity parameters determined for a variety of dysprosium-doped glasses are presented in Table~\ref{JO_table_glass}.
Amongst the ZrF$_4$-based glasses, and even specifically different ZBLAN measurements, there is distinct variation in the values obtained for these parameters.
A contributing factor in this discrepancy is that there is a small magnetic dipole contribution to the \SI{3}{\micro\m} transition which is not considered equally in every analysis~\cite{Cases1991,Schweizer1996,Quimby2017}.
Additionally, the ${^6H_{15/2}\rightarrow^6H_{13/2}}$ transition has appreciably larger reduced matrix elements U$^\lambda$ compared to other ground state transitions resulting in a  disproportionate impact on the fitting anaylsis~\cite{Piramidowicz2008b}.
As a common OH$^{-}$ host impurity absorption coincides spectrally with this transition around \SI{3}{\micro\m},  impurity concentration and how completely its effect is removed from the data naturally varies across investigations.

Measured phenomenological intensity parameters suggest stronger transition intensities in both tellurite (TeO$_2$) and chalcogenide glasses as compared to ZBLAN.
While possessing a larger nominal phonon energy of \SI{650}{\per\cm}, tellurite glass is compelling for dysprosium doping as its transparency window nevertheless extends well beyond \SI{3}{\micro\m}, and is demonstrably superior to ZBLAN in mechanical durability~\cite{Richards2013}.
However, in practice the larger phonon energy severely impacts the $^6H_{13/2}$ lifetime via multiphonon decay and correspondingly has been shown to limit luminescence efficiency to significantly less than 1\%~\cite{Gomes2014}.
This issue is exacerbated by the relative difficulty in limiting OH$^{-}$ impurity in tellurite (and in fact all oxide glasses) as compared to standard ZBLAN.
Chalcogenide on the other hand is not phonon energy limited in the same way, as substantially smaller maximum phonon energy as compared to ZBLAN is common, yet thus far no mid-infrared chalcogenide fiber laser has been demonstrated.
Current limitations are the comparably smaller achievable doping concentration, and propagation losses that are routinely on the order of 1~\si{\decibel\per\m}~\cite{Falconi2017}.
 
\begin{figure}[!htbp]
	\begin{overpic}{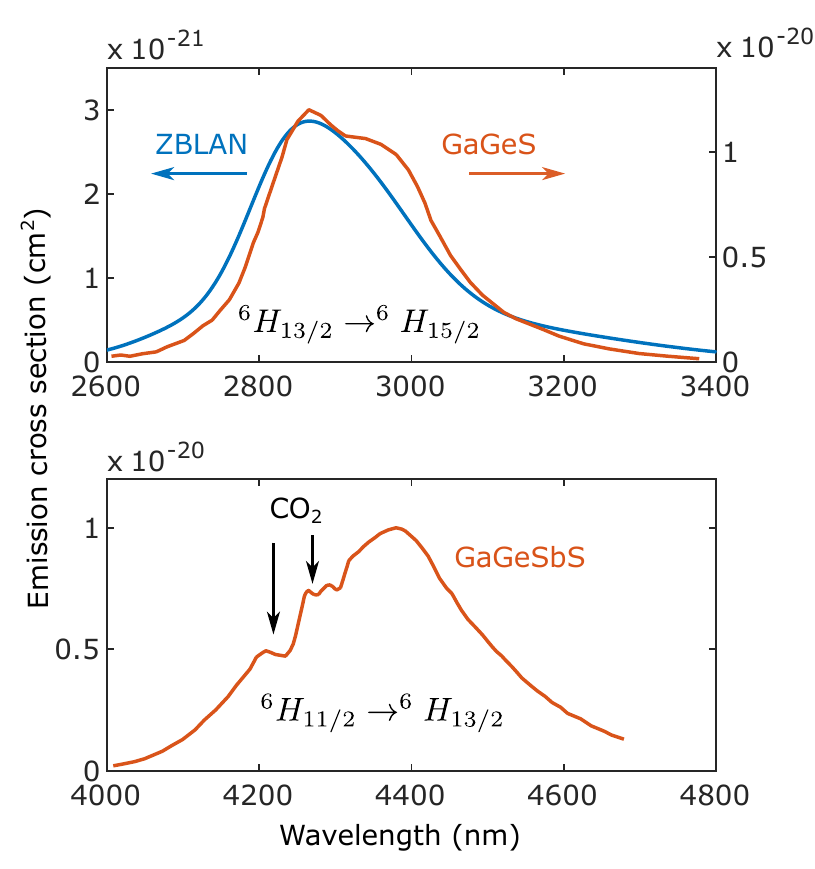}
		\put(15,88){(a)}
		\put(15,40){(b)}
		\end{overpic}
	\caption{Measured emission cross sections: a) \SI{3}{\micro\m} emission cross sections in ZBLAN and a representative chalcogenide glass, GaGeS (data from Refs~\cite{Gomes2010a,Shin1999} respectively) and b) emission cross section of the ${^6H_{11/2}\rightarrow^6H_{13/2}}$ transition in a chalcogenide glass (GaGeSbs). Despite measurement in a nitrogen-purged environment, features arising from residual CO$_2$ absorption remain (data from Refs~\cite{Charpentier2013,Falconi2016}).}
	\label{fig-glass-emission}
\end{figure}
\begin{figure}[!htbp]
	\includegraphics{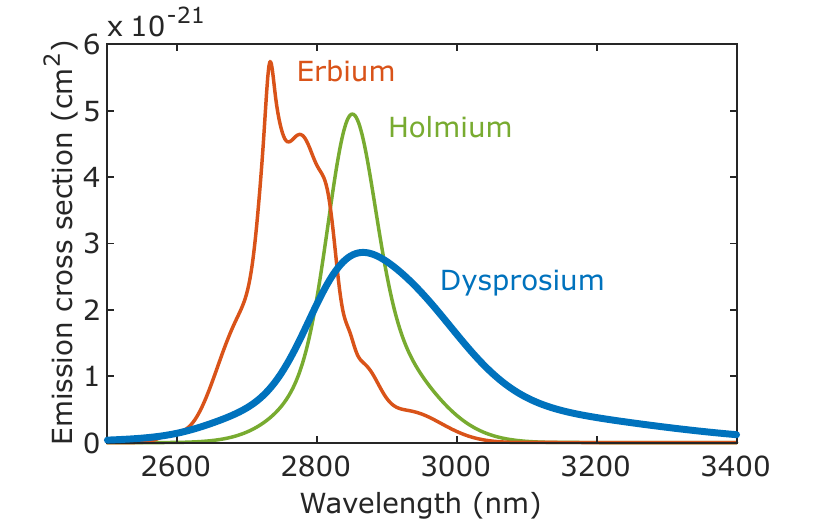}
	\caption{Measured emission cross sections for \SI{3}{\micro\m} transitions in Er, Ho and Dy:ZBLAN. Data for Er and Ho:ZBLAN taken from Refs.~\cite{Li2012d,Sapir2016c} respectively, while Dy:ZBLAN is from~\cite{Gomes2010a}.}
	\label{fig-erhody}
\end{figure}

\begin{table}[!htbp]
	\centering
	\caption{Peak emission cross sections ($\sigma_e$  , \SI{e-21}{\cm^2}) of the low-lying mid-infrared Dy$^{3+}$ transitions in various glasses. Transitions terminate on the the ground ($^6H_{15/2}$) and first excited state ($^6H_{13/2}$), respectively. }
	\label{emission_table_glass}
	\begin{tabular}{@{}cccc@{}}
		\toprule
		Host      & $^6H_{11/2}\rightarrow^6H_{13/2}$ & $^6H_{13/2}\rightarrow^6H_{15/2}$ &      Reference       \\ \midrule
		
		GaLaS     &               11.7                &                9.2                & \cite{Schweizer1996} \\
		GeAsGaSe    &                8.2                &                9.3                &   \cite{Shaw2001}    \\
		GaSbS     &                3.6                &               10.6                &   \cite{Yang2017}    \\
		GeGaSe     &                                   &               12.8                &   \cite{Nemec2000}   \\
		AlF$_3$    &                                   &                 6                 &   \cite{Zhou2016}    \\
		ZBLAN     &                                   &                 3                 &  \cite{Gomes2010a}   \\ \bottomrule
	\end{tabular}
\end{table}
Experimentally measured emission cross sections for the ${^6H_{13/2}\rightarrow^6H_{15/2}}$ transition are shown in Figure~\ref{fig-glass-emission}(a) for both ZBLAN and a representative chalcogenide host, in this case GeGaS. 
Aside from the nominally threefold increase in peak magnitude, there is minimal substantial difference in the spectral shape of emission.
Emission around \SI{3}{\micro\m} in ZBLAN is characterized by a peak value that is nominally 50\% reduced as compared to Ho and Er \SI{3}{\micro\m} transitions in ZBLAN, but as seen in Figure~\ref{fig-erhody} dysprosium exhibits a superior spectral width and infrared coverage. 

Cross section measurement for the  ${^6H_{11/2}\rightarrow^6H_{13/2}}$ transition is seen in Figure~\ref{fig-glass-emission}(b) for a chalcogenide glass host.
As expected, atmospheric CO$_2$ represents a significant effect, whereas even experimental data taken in a gas-purged setup still retain some carbon dioxide artifacts~\cite{Charpentier2013}.
Similarly to measurements in crystalline hosts, numerical values for cross section are obtained via measured fluorescence spectra and the F\"{u}chtbauer-Ladenburg relation, thus residual atmospheric absorption alteration of the measured spectrum manifests as error margin in the calculation. 
Due to the larger phonon energies and corollary optical transparency limitations, emission from the \SI{4}{\micro\m}  dysprosium transition is typically only observed in chalcogenide glasses.
Recently, emission from this transition was observed in an indium fluoride (InF$_3$) based glass fiber which will be discussed in greater detail in section~\ref{4micron_emission}.

Tabulated values for peak cross sections in various hosts are presented in Table.~\ref{emission_table_glass}. 
It is clear that should the practical issues outlined previously be suitably addressed, there are multiple chalcogenide glasses that would be highly desirable for fiber laser applications.
Aside from the chalcogenides, a large increase in emission cross section has also been observed in fluoroaluminate (AlF$_3$-based) glass~\cite{Zhou2016}.
This is an interesting result, as AlF$_3$ can sustain reasonably high rare-earth doping concentrations and exhibits markedly smaller fiber propagation loss in comparison to the investigated chalcogenides. 
In addition, AlF$_3$ glasses are reported to be substantially less hygroscopic than ZrF$_4$ glasses such as ZBLAN, mitigating the introduction of OH$^{-}$ impurity~\cite{Iqbal1991}.
Despite this potential advantage, while record efficiency has been achieved in the visible with Dy:AlF$_3$~\cite{Fujimoto2010,Fujimoto2011}, there has yet to be demonstration a dysprosium doped MIR fluoroaluminate fiber laser.
As a possible means of explanation it should be noted that while impurity losses are theoretically lesser in AlF$_3$ as compared to ZBLAN, intrinsic losses around  \SI{3}{\micro\m} are several times larger.

\begin{table}[!htbp]
	\centering
	\caption{Calculated and experimentally observed excited state lifetimes (ms) of Dy$^{3+}$ in various hosts.}
	\label{lifetime_table_glass}
	\begin{tabular}{@{}cccccc@{}}
		\toprule
		& \multicolumn{2}{c}{$^6H_{11/2}$} & \multicolumn{2}{c}{$^6H_{13/2}$} &                       \\
		Host       & $\tau_{calc}$ &  $\tau_{meas}$   & $\tau_{calc}$ &  $\tau_{meas}$   &       Reference       \\ \midrule
		
		ZBLAN       &     13.7      &      0.0012      &     46.8      &       0.64       &   \cite{Gomes2010a}   \\
		ZBLA       &     13.5      &                  &     51.2      &                  &    \cite{Adam1988}    \\
		GeGaS       &     2.94      &       1.09       &      6.7      &       3.92       &    \cite{Shin1999}    \\
		
		GeAsS       &      2.2      &       0.73       &      5.3      &       4.55       &    \cite{Heo1996}     \\
		GaLaS       &      2.5      &       1.3        &               &                  & \cite{Schweizer1996}  \\
		GeAsGaSe     &      2.4      &        2         &      6.2      &        6         &    \cite{Shaw2001}    \\
		GaSbS       &     2.02      &       1.38       &     4.38      &       3.15       &    \cite{Yang2017}    \\
		TZNF (tellurite) &     4.86      &       0.001      &     15.9      &      0.019       &   \cite{Gomes2014}    \\
		 \bottomrule
	\end{tabular} 
\end{table}
A selection of radiative excited state lifetimes calculated from a Judd-Ofelt analysis in various hosts is presented in Table~\ref{lifetime_table_glass} along with experimentally measured values where available.
From this data the influence of host phonon energy on the measured fluorescence lifetime is immediately clear.
In all of the chalcogenide glasses (both sulfides and selenides) the contribution of non-radiative decay to the observed lifetime is relatively modest for the $^6H_{13/2}$ level.
The multiphonon non-radiative decay rate, $W_{nr}$, can be defined in terms of these lifetimes by the relation $W_{nr}=(1/\tau_{meas})-(1/\tau_{calc})$, where for example in GeAsS this is found to be \SI{106}{\per\second} for $^6H_{13/2}$, resulting in a comparably large luminescence efficiency ($\tau_{meas}/\tau_{rad}$).
Alternatively to empirical measurement, analytical calculation of non-radiative rate between two states separated by energy $\Delta E$ is described by the so-called ``energy-gap law"~\cite{Reisfeld1987,Lume1977}
\begin{equation}
W_{nr} = C[n(T)+1]^{ \Delta E /  \hslash \omega} e^{-\alpha \Delta E}
\end{equation} 
where $n(T)=(e^{\hslash\omega/kT}-1)^{-1}$, $C$, and $\alpha$ are host dependent parameters while $\hslash\omega$  is host phonon energy.
As the separation between $^6H_{11/2}$ and $^6H_{13/2}$ is smaller than that of the first excited and ground states, the non-radiative rates are correspondingly larger, resulting in a more pronounced reduction in observed lifetime.
Non-radiative rates for tellurite are exceedingly high as is directly observed experimentally, with a rate for the \SI{3}{\micro\m} level of \SI{5.2e5}{\per\second}, a primary reason why lasing has not yet been demonstrated in a dysprosium-doped tellurite fiber.

With a phonon energy between that of the chalcogenides and tellurites, ZBLAN ($\hslash\omega=$\SI{600}{\per\cm}) also exhibits strong multiphonon decay for both MIR transition upper states~\cite{Zhu2010b}. 
The lifetime of $^6H_{11/2}$ is dominated by multiphonon decay, which when coupled with the self-terminating nature of this transition and the transparency limitation of ZBLAN beyond \SI{4}{\micro\m}, render MIR lasing from this level likely not feasible.
Despite the low luminescence efficiency, the $^6H_{13/2}$ excited state lifetime however remains long enough in ZBLAN that oscillation thresholds are readily achievable, resulting in multiple demonstrations of Dy:ZBLAN fiber laser systems operating on the ${^6H_{13/2}\rightarrow^6H_{15/2}}$ transition. 
Interestingly, the non-radiative rates observed for Dy:ZBLAN are approximately an order of magnitude greater than that of erbium or holmium for similar separation energies despite the accepted view that non-radiative rates are host, not ion, dependent.
Quimby and Saad have examined this in detail and propose that the lower 5$d$ energy levels of dysprosium as compared to other rare earths plausibly enable a stronger electron-phonon coupling, resulting in the observation of anomalously large rates of non-radiative decay~\cite{Quimby2017}.

\begin{figure}[!htbp]
	\includegraphics{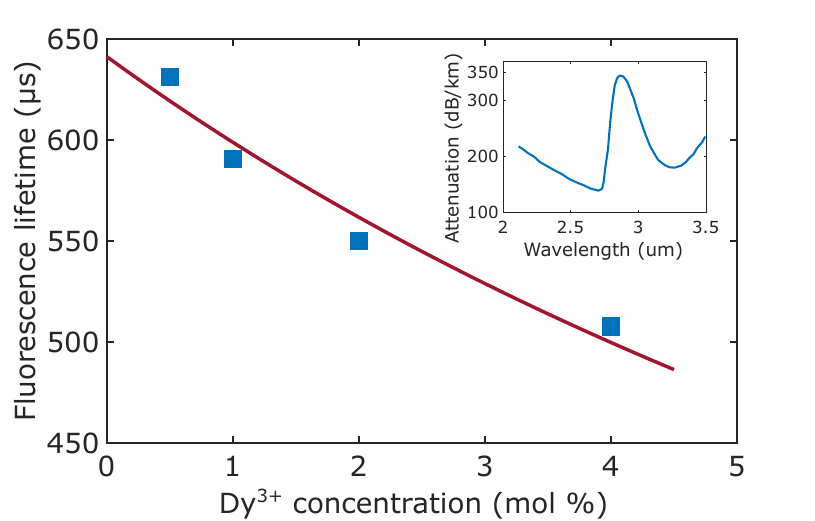}
	\caption{Measurement of impurity-induced fluorescence lifetime concentration quenching of the $^6H_{13/2}$ level in Dy:ZBLAN (data from~\cite{Gomes2010a}). A least squares fit to Eq.~\ref{eq-2} yields intrinsic total lifetime of \SI{641}{\micro\s} and a transfer rate of \SI{110}{\mole\per\percent\per\second}. Inset: attenuation measurement of H$_2$O-doped ZBLAN with pronounced OH$^-$ absorption feature. Data from~\cite{France1984}.}
	\label{fig-quenching-glass}
\end{figure}
Similarly to the observations in crystalline hosts, concentration quenching of the $^6H_{13/2}$ level has been observed in Dy:ZBLAN.
Measurement of fluorescence lifetime as a function of dysprosium concentration is seen in Figure~\ref{fig-quenching-glass}, where moderate decrease is observed at higher ion concentration.
This is attributed to energy transfer from 	$^6H_{13/2}$ to the OH$^{-}$ radical, a well recorded common impurity in fluoride glass (see attenuation data in Figure~\ref{fig-quenching-glass} inset ) which is mitigated by dry environment fabrication, though currently cannot be eliminated entirely.
Gomes et al. \cite{Gomes2010a} assess the energy transfer probability by fitting this measured data with
\begin{equation}
\label{eq-2}
\tau=\left[\frac{1}{\tau_1}+W_t\right]^{-1}
\end{equation}
where $\tau_1$ is the intrinsic total decay time, and $W_t=aN_{conc}$, with $N_{conc}$ the Dy ion concentration and $a$ the transfer probability rate.
The fit produces a value for $a$ of \SI{110}{\mole\per\percent\per\second} which while undesirable is considered to be negligible in comparison to the multiphonon decay rate. 

\subsection{Co-doping}
\begin{figure*}
	\begin{overpic}{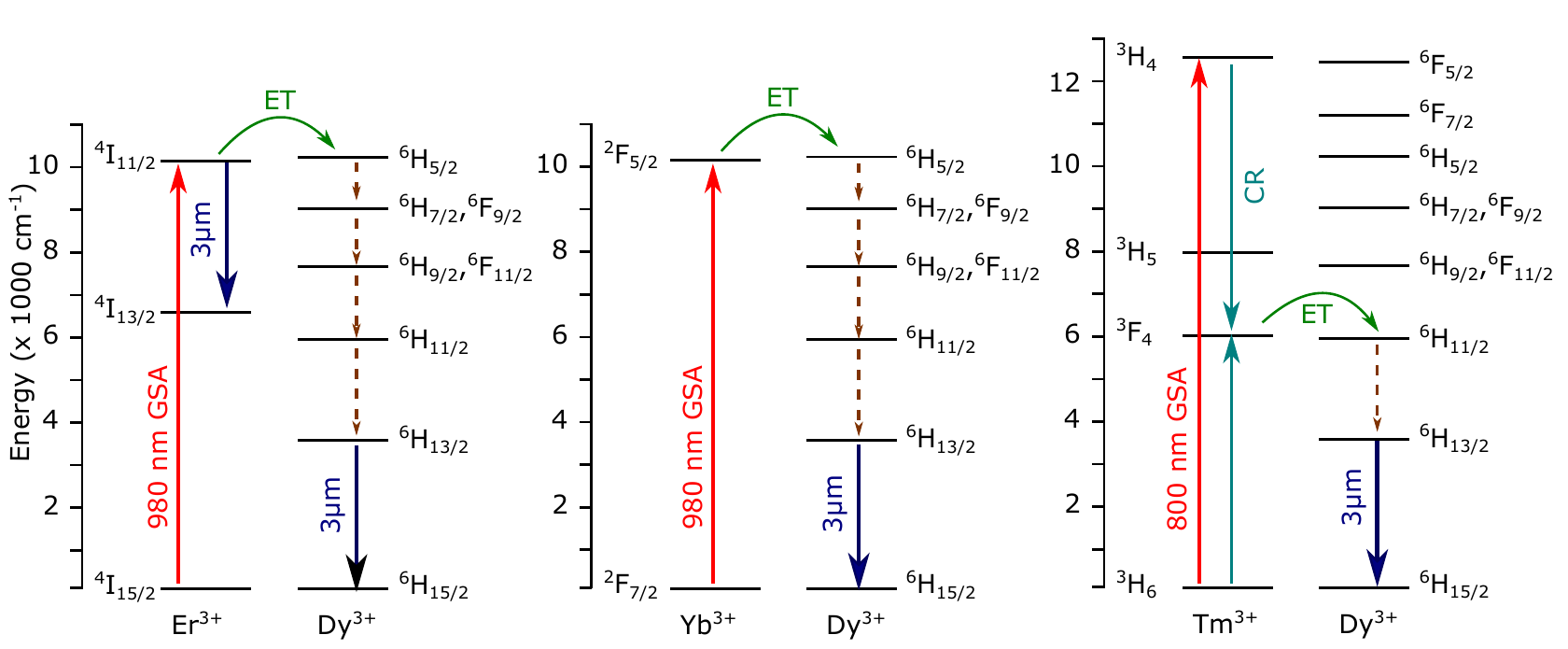}
		\put(0,37){(a)}
		\put(35,37){(b)}
		\put(63,37){(c)}
		\end{overpic}
	\caption{Co-doping schemes for energy transfer (ET) to dysprosium: a) erbium pumped at 980~nm transfers to $^6H_{5/2}$ which populates $^6H_{13/2}$ through successive non-radiative multiphonon relaxation (MPR - dashed arrows), b) ytterbium also pumped at 980~nm, populates $^6H_{13/2}$ through MPR and does not exhibit competing MIR emission, and c) thulium pumped at 800~nm undergoes efficient cross-relaxation (CR) and transfers energy to $^6H_{11/2}$; a single MPR populates $^6H_{13/2}$ for \SI{3}{\micro\m} emission.} 
	\label{fig-codoping}
\end{figure*}
As it is ground state terminated, net-positive gain from the ${^6H_{13/2}\rightarrow^6H_{15/2}}$ transition requires large fractional population inversion, often in excess of 50\% near the emission peak.
To achieve said inversion requires a relatively high pump absorption density, though for efficient laser oscillation this must also be balanced against minimizing signal re-absorption.
A plausible solution is found in a co-doping scheme, where pump light is strongly absorbed by a sensitiser ion which then rapidly transfers energy to a Dy$^{3+}$ acceptor ion.
In this manner dysprosium concentration can be kept low while still potentially experiencing strong inversion, and importantly in a fiber geometry, this opens up the possibility of utilizing cladding pumping.
A classic example of the efficacy of this approach is seen in Er/Yb co-doped fiber, enabling highly efficient and power-scalable diode-pumped double-clad fiber lasers operating on a 3-level transition around \SI{1.55}{\micro\m}.

A co-doped system is most effective when the energy levels of the participating ions are closely resonant with each other.
Phonon-assisted transfer is also possible, but quickly becomes inefficient with increasing mismatch in donor/acceptor energy.
There are several rare-earth candidates for this role, but on initial inspection of Figure~\ref{fig-abs}, it may appear that Nd$^{3+}$ is ideal as the $^6F_{5/2}$ dysprosium level around 800~nm is well matched by the $^4F_{5/2},^2H_{9/2}$ multiplet in neodymium.
The $\sim$800~nm in ground state absorption in neodymium is comparably stronger than that of dysprosium, while high power laser diodes at 808~nm are widely available largely due to the widespread utility of the Nd:YAG laser at \SI{1.064}{\micro\m}.
However, the 808~nm level of Nd is strongly non-radiatively coupled to the upper laser level $^4F_{3/2}$, which makes for quite efficient 4-level lasing, but inefficient transfer to dysprosium ions~\cite{Qi2017}.
More effective alternatives are Yb, Er, and Tm, which as can be seen from Figure~\ref{fig-RE-levels} all possess levels well matched to those of dysprosium: $^2F_{5/2}$ in Yb, and $^4I_{11/2}$ in Er coincide energetically with the $^6H_{5/2}$ dysprosium level, while the $^3H_4$ of Tm overlaps well with Dy $^6H_{11/2}$.

An energy level diagram detailing a potential transfer process between Er$^{3+}$ and Dy$^{3+}$ is shown in Figure~\ref{fig-codoping}(a); ground state absorption around 980~nm in erbium is attractive as high power diodes here are commercially available.
Transfer from the $^4I_{11/2}$ level to $^6H_{5/2}$ subsequently populates the upper dysprosium level for \SI{3}{\micro\m} emission ($^6H_{13/2}$) through effective multiphonon relaxation.
Experimentally, this pathway has been investigated by few authors, though transfer is confirmed by observation of a fluorescence spectrum that is broader than that or Er$^{3+}$ alone due to the contribution from dysprosium emission at longer wavelengths~\cite{Wang2017,Wang2018}.
This work also generates from an extension of Dexter's theory of energy transfer~\cite{Dexter1953,Tarelho1997} an estimate for the microscopic energy transfer constant from donor to acceptor in tellurite glass, $C_{DA}$ of \SI{6.9e-38}{\cm^6\per\s}.
However, as erbium exhibits its own strong \SI{3}{\micro\m} transition (${^4I_{11/2}\rightarrow^4I_{13/2}}$) this co-doping scheme is susceptible to an excited state absorption of the mid-infrared emission.
This perhaps explains the experimental observation that the emitted fluorescence broadens but does not appreciably increase in  intensity.

Alternatively, transfer from Yb$^{3+}$ does not include any excited state absorption processes, while still allowing for the possibility of pumping at the convenient 980~nm wavelength, as seen in the energy level diagram in Figure~\ref{fig-codoping}(b).
Confirmation that this path can be quite efficient in low phonon energy fluoride crystalline hosts is evidenced by the experimental observation of a two order of magnitude reduction in Yb $^2F_{5/2}$ lifetime on the introduction of Dy in a 10:1 Yb/Dy ratio~\cite{Tigreat2001}.
A similar doping ratio in germanate glass resulted in increased MIR fluorescence intensity~\cite{Shen2018}.
A value for the microscopic energy transfer constant of \SI{7.36e-39}{\cm^6\per\s } was obtained in tellurite glass and a transfer efficiency as high as 80\% is calculated based on the reduction in Yb lifetime~\cite{Ye2017}.

Thulium is a particularly interesting candidate as a sensitizer ion and has been investigated in several hosts, both crystals~\cite{Toncelli1999,Tigreat2001,Hu2019,Doualan2018} and glasses~\cite{Li2011,Ma2015,Tian2012,Tian2012a,Wang2017,Heo1997}.
While there is a degree of overlap between the energies of $^3H_4$ in Tm and $^6F_{3/2}$ in Dy, there is a general consensus thus far that the dominant transfer mechanism is as illustrated in Figure~\ref{fig-codoping}(c).
Pumping around 790~nm produces excited population of the $^3H_4$ level in Tm, which then undergoes rapid and highly efficient cross-relaxation with the ground state, resulting in $^3F_4$ population.
This then resonantly transfers energy to $^6H_{11/2}$ in Dy, which is non-radiatively coupled to the upper state of the \SI{3}{\micro\m} transition.
This view is supported by experimental observation of the lifetimes of both involved thulium excited states, wherein the presence of dysprosium ions only has observable effect on the lifetime of $^3F_4$.
The microscopic energy transfer parameter calculated for representative fluorophosphate and tellurite glasses are \SI{1.63e-38}{\cm^6\per\s} and \SI{1.48e-38}{\cm^6\per\s} respectively~\cite{Tian2012,Ma2015}.
While smaller than the value determined for Er co-doping in tellurite glass, it is important to note that in the thulium scheme possible excited state absorptions are eliminated and overall quantum efficiency is effectively doubled by the cross-relaxation mechanism. 

There remains considerable scope for continued investigation into identifying and characterizing co-doping schemes.
For example, co-doping with a view towards development of 4~\si{\micro\m} sources has yet to be considered, while amongst the proposed transfer mechanisms for 3~\si{\micro\m} emission, parametric optimization of dopant ratios is distinctly lacking.
Furthermore, none of the proposed schemes has yet been utilized to demonstrate laser emission in any host.

\section{Laser Emission}
\label{sec:lasers}
In the context of spectroscopic study, dysprosium represents a small fraction of the research effort devoted to optically active rare-earth ions.
This disparity becomes even more apparent when experimental demonstration of laser emission is considered, to such an extent that rather than largely representative, this review is a complete collection to the best of our knowledge of MIR laser results reported to date.
Despite this relatively small body of work, it is striking that the frequency of published dysprosium mid-infrared laser results has dramatically increased in recent years. 
 
\begin{table*}
	\centering
	\caption{Performance parameters of dysprosium-doped crystalline lasers.}
	\label{laser_crystal_table}
\begin{tabular}{@{}cccccc@{}}
	\toprule
	       Host         & $\lambda_{pump}$~(\si{\micro\meter}) & $\lambda_{laser}$~(\si{\micro\meter}) & Slope efficiency~$\eta$ (\%) & Max. output energy &      Year, Reference       \\ \midrule
	   BaY$_2$F$_8$     &              flashlamp               &                 3.022                 &                              &                    &  1973~\cite{Johnson1973b}  \\
	      LaF$_3$       &                1.064                 &                 2.97                  &              9               &       500~mJ       & 1980~\cite{Antipenko1980a} \\
	        
	   BaYb$_2$F$_8$    &                 1.3                  &                  3.4                  &              4               &       70~mJ        &    1997~\cite{Djeu1997}    \\
	   YLF         &                 1.73                 &                 4.34                  &              5               &       0.2~mJ       &   1991~\cite{Barnes1991}   \\
	   CaGa$_2$S$_4$    &                 1.3                  &                 4.38                  &             1.6              &      0.12~mJ       &  1999~\cite{Nostrand1999}  \\
	PbGa$_2$S$_4$ (PGS) &                 1.3                  &                  4.3                  &              1               &      0.35~mJ       & 2009~\cite{Doroshenko2009} \\
	        PGS         &                 1.3                  &                 4.29                  &              3               &      0.09~mJ       &    2010~\cite{Jan2010}     \\
	        PGS         &                 1.7                  &                 4.36                  &              8               &        7~mJ        & 2011~\cite{Jelinkova2011}  \\
	        PGS         &                 1.7                  &                  4.3                  &              8               &     67~mW (CW)     & 2013~\cite{Jelinkova2013a} \\ \bottomrule
\end{tabular} 
\end{table*}
	\subsection{Crystalline lasers}
Despite the notably fewer recently published results as compared to fiber, dysprosium crystalline lasers are of unique importance for multiple reasons, not least of which is a historical perspective.
Lasing and key early spectroscopic data from a crystal host predates the first Dy$^{3+}$ fiber demonstration by four decades, laying the necessary foundation for all subsequent work.
Further, attainable pulse energies in bulk dysprosium crystalline lasers well exceed what is currently possible in a fiber geometry.
Perhaps the most important distinction however is the availability of high quality crystals with exceptionally low phonon energies, enabling the only demonstrations to date of lasing beyond \SI{4}{\micro\m} from dysprosium. 
Indeed no fiber laser has yet been demonstrated beyond this critical benchmark.
Key performance metrics of the Dy:crystal lasers demonstrated to date are summarized in Table~\ref{laser_crystal_table}.
	
	\subsubsection{Laser emission from the $^6H_{13/2}\rightarrow^6H_{15/2}$ transition}
The first reported dysprosium laser emission originated from a BaY$_{2}$F$_{8}$ crystal~\cite{Johnson1973b} operating on the 3~\si{\micro\m} ${^6H_{13/2}\rightarrow^6H_{15/2}}$ transition.
Despite the relatively low phonon energy of this crystal (\textless500~\si{\per\cm}) at room temperature, immersion of the crystal in liquid nitrogen was necessary to extend the upper state lifetime and achieve oscillation threshold.
Pump radiation was supplied by a high-energy xenon flash lamp, and though not reported, the broadband nature of the pump source presumably did not produce a high efficiency system.
Emergence of Nd:YAG lasers facilitated the few further developments of 3~\si{\micro\m} Dy:crystal lasers enabling pump absorption targeted specifically at the $^6H_{7/2},^6F_{9/2}$ and $^6H_{9/2},^6F_{11/2}$ multiplets at 1.1 and \SI{1.3}{\micro\m} respectively, and operation at room temperature.
While conversion efficiencies of these subsequent systems were less than 30\% of the Stokes limit, lasing wavelengths were demonstrated both near the peak of the emission cross section and well in the long wavelength tail at \SI{3.4}{\micro\m}, clearly demonstrating the utility of the Dy ion as a broadband MIR laser source~\cite{Antipenko1980a,Djeu1997}.

	\subsubsection{Laser emission from the $^6H_{11/2}\rightarrow^6H_{13/2}$ transition}
In terms of published research results, interest in 3~\si{\micro\m} dysprosium crystal lasers has been wholly replaced more recently by the development of lasers emitting around \SI{4.3}{\micro\m} on the ${^6H_{11/2}\rightarrow^6H_{13/2}}$ transition.
This shift is in part explained by the desire to replace optical parametric oscillator systems with directly lasing solid-state bulk crystalline laser penetrating further into the MIR for applications such as counter-measures and remote sensing.
As discussed, this dysprosium transition is uniquely suited to meet this need as it is situated close to the ground state (small quantum defect limit) and emits near a strong carbon dioxide absorption feature.
The first demonstration of a 4~\si{\micro\m} Dy crystal laser attempted to take advantage of this favorable energy level position by pumping the $^6H_{11/2}$ upper laser level directly with an Er:YLF Q-switched laser emitting at \SI{1.73}{\micro\m}.
Both the efficiency and maximum output pulse energy realized fall short of the performance of 3~\si{\micro\m} systems, attributed mainly to the self-terminating nature of the transition, and of chief importance, the low optical quality of the crystal~\cite{Barnes1991}.

The relative purity of host materials becomes an increasingly significant concern in all laser systems as the emission wavelengths extend further into the MIR where many common impurities such as water exhibit strong absorption.
As fluoride and chloride crystals tend to be  deliquescent, further development of dysprosium 4~\si{\micro\m} lasers alternatively looked to sulfide crystals.
Early work with CaGa$_2$S$_4$ produced still poorer performance than the initial laser demonstration, owing to the larger quantum defect of pumping with a Nd:YAG laser operating at \SI{1.3}{\micro\m} and lingering issues of crystal quality~\cite{Nostrand1999}.
  
\begin{figure}
	\includegraphics{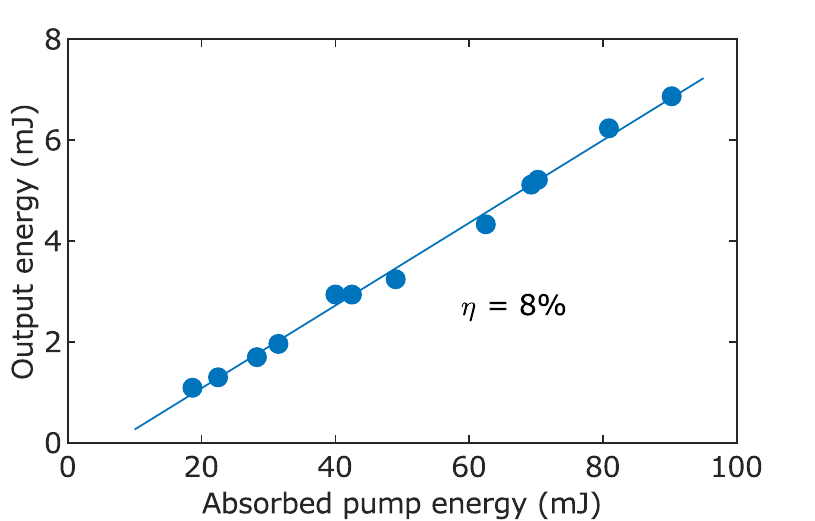}
	\caption{Output pulse energy of a Dy$^{3+}$:PGS \SI{4.3}{\micro\m} laser as a function of absorbed \SI{1.7}{\micro\m} pump energy. Both maximum output energy (7~mJ) and slope efficiency ($\eta$) represent current class-leading performance. Data taken from Ref.~\cite{Jelinkova2011}.}
	\label{fig-PGS-output}
\end{figure}

A relatively successful host is identified in PbGa$_2$$S_4$, or PGS, which possesses lower phonon energy in comparison to CaGa$_2$S$_4$ while retaining resistance to moisture attack, and exhibits a six-fold increase in Dy ion solubility~\cite{Doroshenko2009}.
Record output energy is then achieved with this host when pumped at \SI{1.3}{\micro\m} with a flashlamp-pumped Nd:YAG laser.
Due to the self-terminating nature of this transition, rapid population of the upper laser level is critical to achieving substantial output, and an effort to replace the flashlamp-pumped Nd:YAG laser with a diode-pumped system proved unable to maintain required pulse energies, resulting in reduced output pulse energy~\cite{Jan2010}.
A sizable improvement to current record performance is achieved in a PGS host when returning to the resonant pumping scheme using an Er:YLF laser as the pump source.
As seen in the data presented in Figure~\ref{fig-PGS-output}, efficiency is increased 60\% relative to the previous record, and output energies increase an order of magnitude to the mJ level~\cite{Jelinkova2011}.
 
\begin{figure}
	\includegraphics{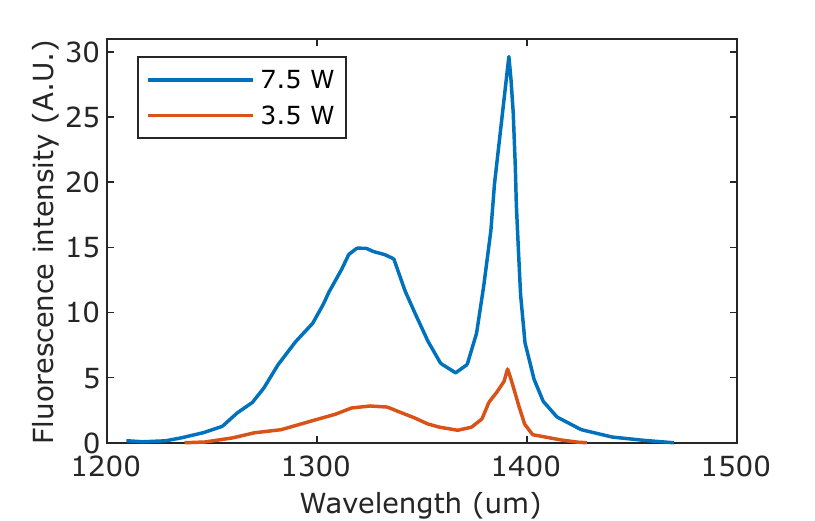}
	\caption{Fluorescence emission from Dy:PGS originating from \SI{1.7}{\micro\m} pump ESA. This mechanism is likely responsible for the demonstration of pure CW \SI{4}{\micro\m} laser operation. Data taken from Ref~\cite{Jelinkova2013a}.}
	\label{fig-CW-4micron-ESA}
\end{figure}

Commercial availability of diode lasers emitting at \SI{1.7}{\micro\m} has enabled the most recent success in Dy:PGS 4~\si{\micro\m} lasers, with the demonstration of pure CW operation, while maintaining the record efficiency previously demonstrated~\cite{Jelinkova2013a}.
This result is somewhat unanticipated as the transition is self-terminating, typically prohibiting pure CW oscillation. 
The authors proposed that excited state absorption (ESA) of pump radiation from the lower laser level provides the required de-population for CW oscillation.
Absorption of a \SI{1.7}{\micro\m} photon from $^6H_{13/2}$ should result in population of $^6H_{7/2}$ and produce some re-emission around 1.1~\si{\micro\m}.
However even in PGS this level is strongly non-radiatively coupled to next lower energy state, $^6H_{9/2}$, from which fluorescence at 1.3~\si{\micro\m} is observed (see Figure~\ref{fig-CW-4micron-ESA}).
A near quadratic dependence on pump power is noted as evidence of a two-photon pump ESA process, contributing to the CW operation~\cite{Jelinkova2013a}.

	\subsection{Glass fiber lasers}
\begin{table*}
	\centering
	\caption{Performance parameters of CW dysprosium-doped glass fiber lasers.}
	\label{laser_glass_table}	
\begin{tabular}{cccccc}
	\toprule
	 Host   & $\lambda_{pump}$~(\si{\micro\meter}) & $\lambda_{laser}$~(\si{\micro\meter}) & Slope efficiency~$\eta$ (\%) & Max. output power~(mW) & Year, Reference \\ \midrule
	 ZBLAN  &                 1.1                  &                  2.9                  &             4.5              &          275           &      2003~\cite{Jackson2003}       \\
	        &                 1.3                  &                 2.96                  &             19.6             &          180           &      2006~\cite{Tsang2006a}       \\
	        &                 1.1                  &                 2.95                  &              23              &           90           &      2011~\cite{Tsang2011}       \\
	        &                 2.8                  &              3.04 / 3.26              &           51 / 37            &           80           &      2016~\cite{Majewski2016c}       \\
	        &                 2.8                  &         2.95 - 3.35 (tunable)         &                              &           30           &      2016~\cite{Majewski2016}       \\
	        &                 1.1                  &                 2.98                  &              18              &          554           &      2018~\cite{Sojka2018a}       \\
	        &                 1.7                  &         2.81 - 3.38 (tunable)         &              21              &          170           &      2018~\cite{Majewski2018a}       \\
	        &                 2.8                  &                 3.15                  &              73              &          1060          &      2018~\cite{Woodward2018b}       \\
	        &                 2.8                  &                 3.24                  &              58              &         10.1~W         &      2019~\cite{Fortin2019}       \\
	InF$_3$ &                 1.7                  &                 2.945                 &              14              &           60           &      2018~\cite{Majewski2018b}       \\ \bottomrule
\end{tabular}
\end{table*} 

Dysprosium-doped glass fiber lasers are comparatively to bulk lasers, a more recent technology, though are drawing  substantially greater research interest and effort of late.
Currently, the number of reported laser results in a fiber geometry well exceeds those of solid-state crystals, and notably the majority of these have only been in the past few years.	
This is to a certain extent driven by the inherent advantages of a fiber format over that of a bulk crystal, such as diffraction-limited beam quality and long gain lengths.
Of course MIR dysprosium fibers require a suitable glass host which has low phonon energy and can be made and drawn into fiber.
Though we have discussed several possibilities above, dysprosium fiber lasers and indeed the entire field of MIR fiber lasers thus far has been thoroughly dominated by ZBLAN.
A comprehensive selection of Dy laser results are presented in Table~\ref{laser_glass_table}, with only a single instance employing a glass other than ZBLAN.

\subsubsection{CW Dy:ZBLAN fiber lasers}
\begin{figure}[!htbp]
	\includegraphics{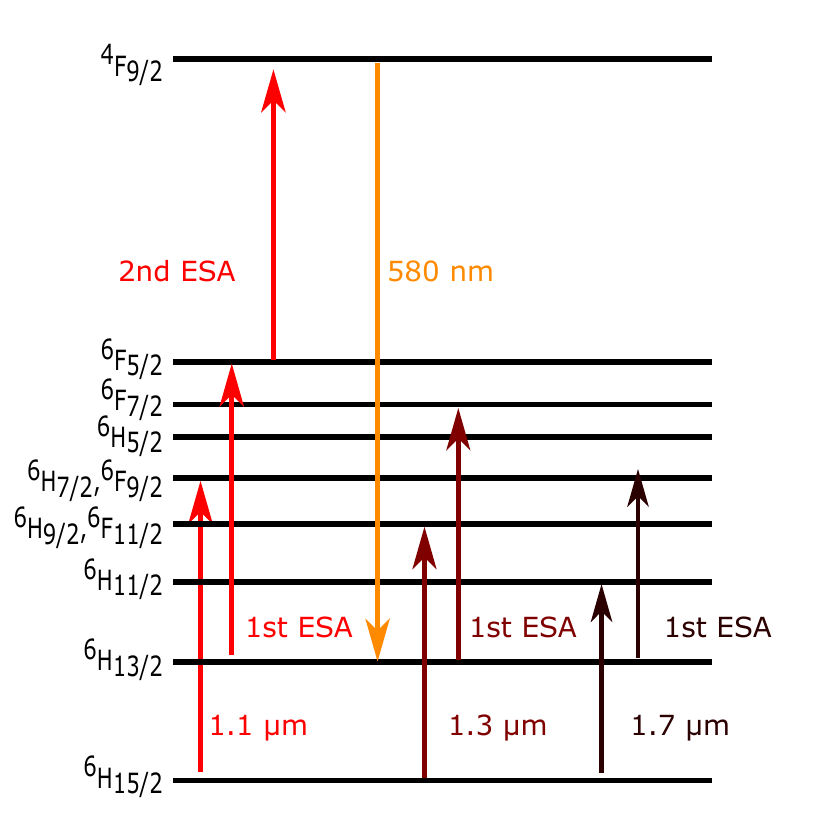}
	\caption{Dy$^3+$ energy level diagram indicating near infrared pump excited state absorption processes.}
	\label{fig-pump-ESA}
\end{figure}

The first realization of a dysprosium doped ZBLAN fiber laser utilized a Yb fiber laser operating at \SI{1.1}{\micro\m} pumping the core of a singly clad fiber~\cite{Jackson2003}.
A fairly large value of cavity loss due largely to the use of a perpendicular cleave as the output coupler resulted in a large value of injected power threshold of 1.8~W, however the availability of pump power allowed output power up to 275~mW, a value not exceeded until very recently.
The efficiency of this system was quite low, at 4.5\%.
This represents only 10\% of the Stokes limit, attributed mainly to the deleterious effect of pump excited state absorption (ESA), which is a recurring issue in many near-infrared pumped dysprosium fiber laser systems.
Mechanisms for pump ESA are illustrated in Figure~\ref{fig-pump-ESA}; when pumping at 1.1~\si{\micro\m} absorption of a pump photon from $^6H_{13/2}$ populates $^6F_{5/2}$ which has a small lifetime, but under strong pumping conditions an additional absorption from this level leads to population of the metastable $^4F_{9/2}$ level.
Emission from the ${^4F_{9/2}\rightarrow^6H_{13/2}}$ transition produces visible yellow fluorescence, the experimental observation of which in this original work confirmed the presence of pump ESA.

To reduce the impact of ESA on laser performance, pumping at \SI{1.3}{\micro\m} has also been demonstrated using a Nd:YAG laser~\cite{Sennaroglu2006}.
As seen in Figure~\ref{fig-pump-ESA}, there is one quasi-resonant excited state absorption from $^6H_{13/2}$ when pumping with \SI{1.3}{\micro\m} photons, in contrast to the two transitions observed in the initial Dy:ZBLAN result, thus ESA is still possible, but potentially mitigated to a degree.
Though the efficiency was still well below the Stokes limit, this pumping scheme achieves a four-fold efficiency increase.
An additional factor in the performance improvement was the implementation of a higher Q cavity employing a butt-coupled 50\% reflective mirror as the output coupler in place of a cleaved facet.
A study of output coupler optimization in a \SI{1.1}{\micro\m}	pumped system yielded further improvement in efficiency to 23\%, indicating that proper cavity design can compensate somewhat the effect of ESA through reduced threshold pump intensities~\cite{Tsang2011}.

Further reduction in quantum defect is found when pumping the $^6H_{11/2}$ level at 1.7~\si{\micro\m}, which in ZBLAN is completely quenched by multiphonon relaxation, effectively populating the upper laser level.
One complicating factor in this scheme however is availability of high brightness laser sources. 
Raman fiber lasers are a compelling solution here, and served as the pump source in the lone demonstration of Dy:ZBLAN pumped at 1.7~\si{\micro\m}~\cite{Majewski2018a}.
Unfortunately, the efficiency of this system fell well short of the Stokes limit, appearing to also be negatively impacted by pump ESA, the specific mechanism of which is illustrated in Figure~\ref{fig-pump-ESA}.
Through numerical modeling of pump absorption, a preliminary estimate for this ESA cross section is provided as \SI{1e-25}{\m^{-2}} though the specific spectral character is as yet undetermined.

\begin{figure}[!htbp]
	\includegraphics{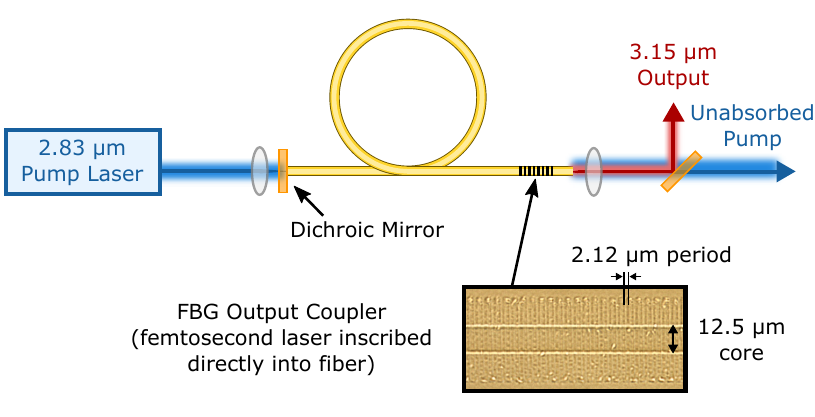}
	\caption{Experimental layout of a high efficiency in-band-pumped Dy:ZBLAN fiber laser using a direct inscription FBG as the output coupler. Reproduced with permission~\cite{Woodward2018b}. 2018, OSA.}
	\label{fig-watt-level}
\end{figure}

\begin{figure}[!htbp]
	\includegraphics{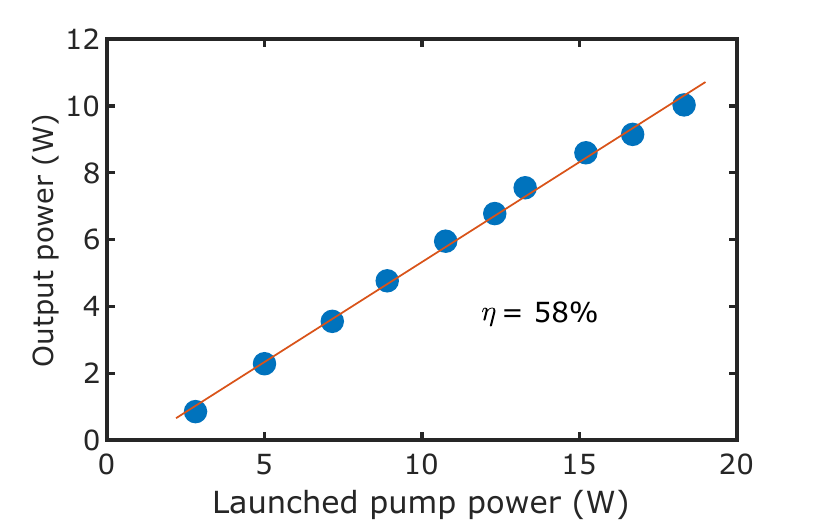}
	\caption{Output power of a \SI{3.24}{\micro\m} monolithic Dy:ZBLAN fiber laser as a function of launched \SI{2.83}{\micro\m} pump power. The maximum output power achieved is 10.1~W while the average slope efficiency ($\eta$) is 58\%. Data taken from Ref.~\cite{Fortin2019}.}
	\label{fig-laval-10W}
\end{figure}

A simple resonant or in-band pumping scheme allows for fully exploiting the inherent efficiency of the ${^6H_{13/2}\rightarrow^6H_{15/2}}$ transition.
While this requires typically an additional MIR fiber laser as the pump source, the performance and relative maturity of diode pumped Er:ZBLAN make this an ideal candidate~\cite{Fortin2015,Aydn2018}.
The first demonstration of an in-band pumped Dy system readily achieved a record efficiency for 3~\si{\micro\m} class fiber lasers of 51\%~\cite{Majewski2016c}.
Additionally, due to the pumping scheme  red-shifting the gain peak, the free running wavelength of 3.04~\si{\micro\m} represented the first Dy:ZBLAN fiber laser operating beyond 3~\si{\micro\m}.
As a quasi-three level system, further extension of the emission wavelength in free running operation was possible through length tuning, where a longer fiber length produced emission at 3.26~\si{\micro\m} though with the necessary accompaniment of efficiency reduction due to increases in both re-absorption and background loss.
Of note is that this system required free-space coupling of the pump beam into the Dy fiber, with then a cavity defined by butt-coupled mirrors.
A clear refinement of this experimental approach in the form of fiber Bragg gratings (FBGs) has recently produced substantial improvement in output power and conversion efficiency.
Replacing the bulk optic output coupler with a partially reflective FBG as seen in the experimental layout Figure~\ref{fig-watt-level} led to current record efficiency of 73\% and maximum output power above the 1~W level~\cite{Woodward2018b}.
An order of magnitude improvement in maximum output power is achieved when free space optics are removed from the system completely.
Recent demonstration of a monolithic system in which the pump laser is directly spliced to the Dy:ZBLAN fiber and all cavities are closed by FBGs exceeded 10~W of output power (see Figure~\ref{fig-laval-10W}) for the first time while retaining a reasonably high efficiency that is nominally 70\% of the fundamental Stokes limit~\cite{Fortin2019}.

\begin{figure}[!htbp]
	\includegraphics{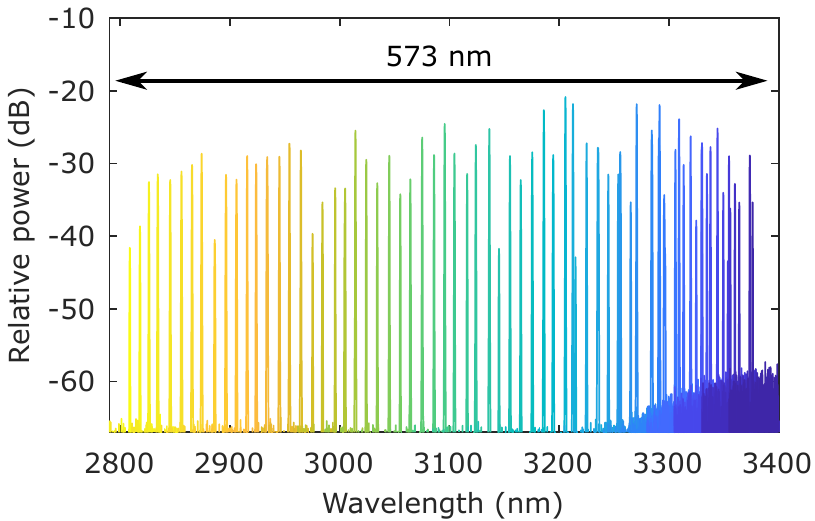}
\caption{Tuning range of the 1.7~\si{\micro\m}-pumped Dy:ZBLAN fiber laser demonstrated in Ref.~\cite{Majewski2018a}.}
	\label{fig-tuning-range}
\end{figure}

While effective in the achievement of high power and efficiency, FBG-based systems enforce a limit on wavelength tunability. 
Considering the broad nature of dysprosium emission, covering the absorption features of multiple functional groups (e.g. OH, NH, and CH), the prospect of continuously tunable Dy:ZBLAN lasers have clear utility.
An extended cavity arrangement incorporating a bulk diffraction grating in an in-band pumped system proved to be continuously tunable (grating angle adjustment) over a range of 400~nm~\cite{Majewski2016}.
The lower wavelength limit for this range of 2.95~\si{\micro\m} is defined in small measure by the particular optics, but more fundamentally by the in-band pump scheme, as gain is quickly reduced at wavelengths approaching that of the pump.
Access to larger portions of the intrinsic bandwidth of the transition is accomplished in a scheme pumped at 1.7~\si{\micro\m} with a Raman fiber laser~\cite{Majewski2018a}.
As seen in the data presented in Figure~\ref{fig-tuning-range}, narrowband laser emission is achieved over a range of nearly 600~nm, representing the most widely tunable rare-earth doped laser to date, and uniquely bridging the spectral gap between Er and Ho systems around 2.9~\si{\micro\m}~\cite{Libatique2000,Crawford2015} and those operating on the Er 3.5~\si{\micro\m} transition~\cite{Sapir2016c}.

\begin{figure*}
	\begin{overpic}{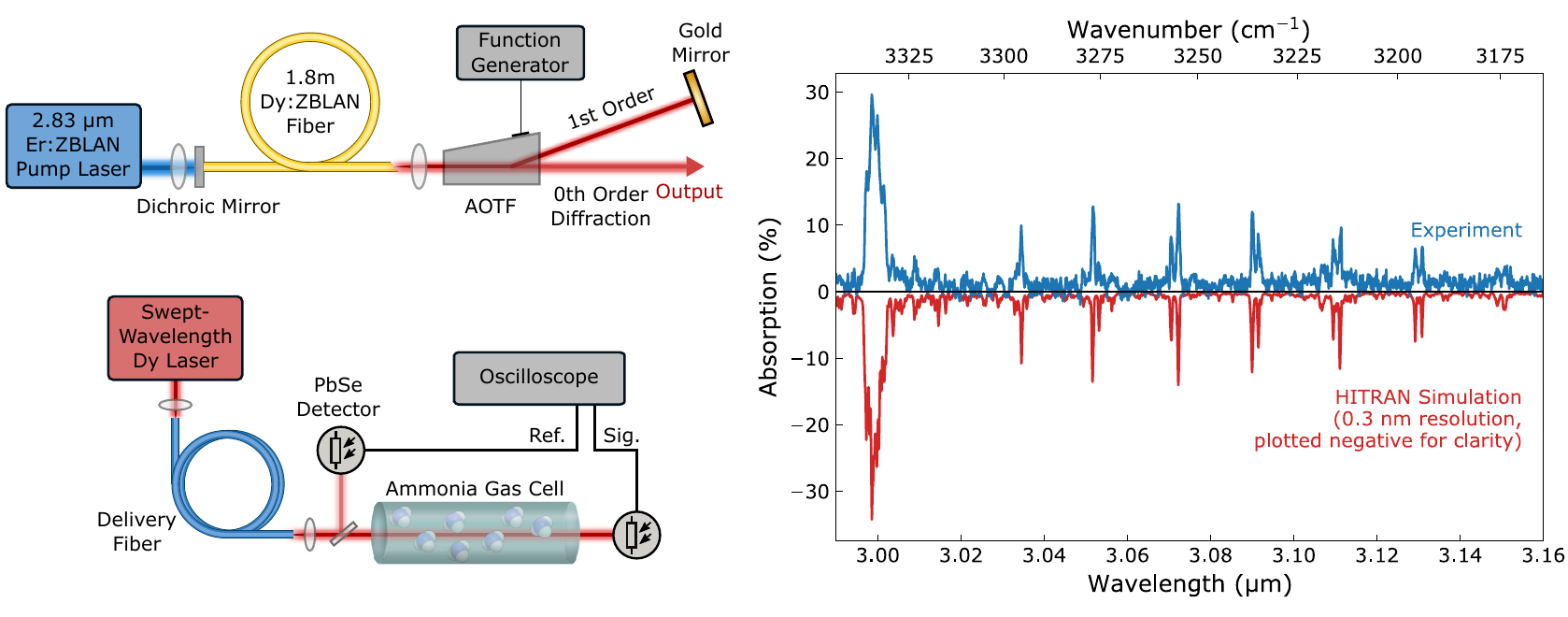}
	\put(0,37){(a)}
	\put(0,22){(b)}
	\put(49,37){(c)}
	\end{overpic}
	\caption{Swept wavelength Dy:ZBLAN fiber laser and ammonia (NH$_3$) real-time gas sensing as demonstrated in Ref.~\cite{Woodward2018a} ; a) layout of the swept wavelength source based on AOTF tuning, b) balanced detection gas sensing setup, c) processed data mapping time to wavelength clearly resolving mutliple ro-vibrational NH$_3$ features.}
	\label{fig-AOTF-NH3}
\end{figure*}

Wavelength sweeping via angle tuning of a bulk diffraction grating can be electronically controlled by a motorized rotation stage for example, yet a substantial improvement in sweep speed is found alternatively with the use of an acousto-optic tunable filter (AOTF)~\cite{Woodward2018a}.
The first order diffraction from an AOTF is narrowband with a center frequency determined by the applied radio frequency (RF) drive signal to the acousto-optic crystal.
Placing this element in the extended cavity configuration shown in Figure~\ref{fig-AOTF-NH3}(a) allowed for rapid electronic tuning of Dy emission wavelength over 360~nm range from 2.89 to 3.25~\si{\micro\m} in a time span on the order of 50~ms.
Such rapid swept-wavelength operation makes this source promising for real time gas sensing of species exhibiting absorption in this wavelength range, ammonia (NH$_3$) being a particularly relevant candidate.
Experimental implementation of a straightforward balanced detection scheme as seen in Figure~\ref{fig-AOTF-NH3}(b) illustrates the utility of this approach to gas sensing, as the resulting data (Figure~\ref{fig-AOTF-NH3}(c)) clearly resolves multiple ro-vibrational features of this molecule, as confirmed by the accompanying HITRAN database simulation. 

\subsubsection{Pulsed Dy:ZBLAN fiber lasers}

Although the majority of Dy:ZBLAN work to date has focused on the CW operation regime, access to a wider range of potential application necessarily requires pulsed laser source as well.
As particular examples, differential absorption lidar and industrial polymer processing both may benefit from high peak power or energy pulses with a large degree of molecular absorption discrimination in the MIR.
In efforts to meet this need, recent significant work in pulsed Dy:ZBLAN fiber lasers has seen rapid progression, covering pulse durations from the nanosecond down to true `ultrafast' picosecond scales.
A collection of performace parameters for these pulsed dysprosium fiber systems is presented in Table.~\ref{pulse-table}.
\begin{table*}
	\centering
	\caption{Performance of temporally modulated Dy fiber lasers.}
	\label{pulse-table}	
	\begin{tabular}{cccccc}
		\toprule
		Mode of operation   & Repetion rate & Pulse duration &    Pulse energy     &  Wavelength (\si{\micro\m}) &    Year, Reference      \\ \midrule
		Q-switched (active)  & 0.1 - 20~kHz  &     270~ns     &  \SI{12}{\micro\J} &2.97 - 3.23 & 2019 \cite{Woodward2019}  \\
		Q-switched (passive) &  47 - 86~kHz  &     740~ns     &   1~\si{\micro\J} & 3.04  & 2019 \cite{Woodward2019}  \\
		Q-switched (passive) &   166.8~kHz   &     795~ns     & \SI{1.51}{\micro\J}&2.71 - 3.08 &  2019 \cite{Luo2019b}   \\
		Mode-locked (FSF)   &   44.5~MHz    &     33~ps      &       2.7~nJ   & 2.97 - 3.3     & 2019 \cite{Woodward2018} \\
		Mode-locked (NPE)&60~MHz&828~fs&4.8~nJ&3.1& 2019 \cite{Wang2019}\\
		Swept-wavelength& &rapidly varying CW wavelength&&2.89 - 3.25 & 2019 \cite{Woodward2018a}\\
		
		\bottomrule
	\end{tabular}
\end{table*}

\paragraph{Q-switching}
Active Q-switching proves a logical extension of the AOTF arrangement shown in Figure~\ref{fig-AOTF-NH3}(a) as in place of sweeping the RF drive frequency, a simple on/off gating produces Q-switched pulses on the nanosecond scale~\cite{Woodward2019}.
Modulation parameters and inversion dynamics can define regions of multiple pulse operation of a Q-switched laser system, which is inhibited in this system experimentally by limiting the modulator `on' time to 20~\si{\micro\s}, resulting in stable single pulses of duration as short as 270~ns with energy up to 12~\si{\micro\joule}.
Due to the intrinsic physical properties of the AOTF, rise time of the diffraction is substantially slower than for example a typical acousto-optic modulator (AOM), i.e. $\tau_{rise}=$~\SI{25}{\micro\s} in this work, whereas an AOM is typically \textless1~\si{\micro\s}.
Detailed numerical modeling confirms this rise time to substantially limit attainable pulse energy and duration, indicating that there remains scope for continued development of actively Q-switched Dy:ZBLAN fiber lasers.

Though active Q-switching generally produces higher energy pulses, passive Q-switching with saturable absorber materials is often desirable due to smaller footprints and substantial reduction in system complexity as drive/control electronics are unnecessary.
In the near-infrared, saturable absorber mirrors (SESAMs) are the common choice for this application, though due to band edge limitations, commercial availability of semiconductor saturable absorbers (SAs) operating in the MIR is exceedingly limited at present.
Thus there is significant recent research effort investigating alternative saturable absorber materials, such as 2D nanomaterials.
With a slight modification to the cavity in Figure~\ref{fig-AOTF-NH3}(a) wherein the AOTF section is replaced an SA arrangement consisting of a tight focusing lens and a silver mirror coated with black phosphorous (BP), passive Q-switching has been successfully demonstrated from a Dy:ZBLAN fiber laser~\cite{Woodward2019}.
This simple arrangement generates stable single pulse operation with energy up to 1~\si{\micro\joule} and duration of 740~ns.
A modest improvement in performance is recently achieved utilizing PbS nanoparticles as the saturable absorber, achieving an increase in pulse energy to 1.5~\si{\micro\joule} at a duration of 795~ns~\cite{Luo2019b}.
Of note however is that this particular cavity employed a bulk diffraction grating, and additionally is pumped by a Yb fiber laser in contrast to in-band pumped systems.
These alterations result in a continuously tunable system over a range of 370~nm, and impressively due to the high power pump source, a lower wavelength limit of 2.71~\si{\micro\m}, which is a notable blue-shift to the limitations of previous tunable systems.

\begin{figure}[!htbp]
	\includegraphics{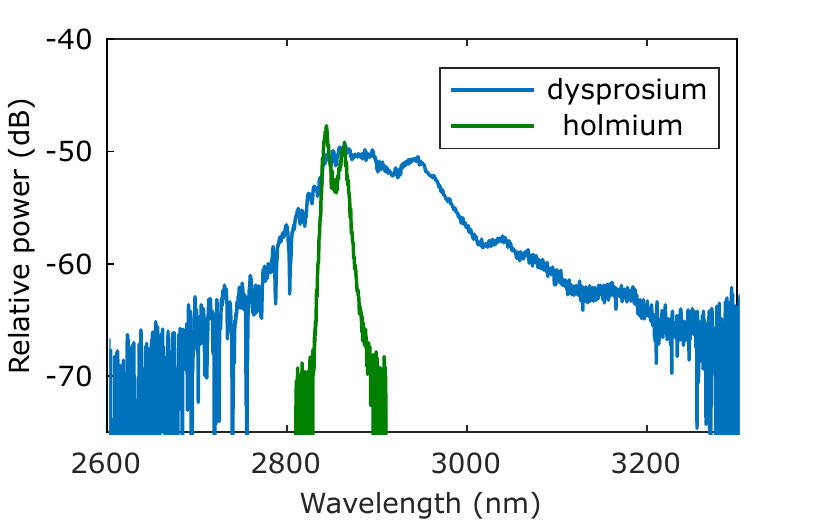}
	\caption{Experimentally measured amplified spontaneous emission (ASE) spectra of Ho:ZBLAN and Dy:ZBLAN fibers. Data from Ref.~\cite{Majewski2018a}.}
	\label{fig-HoDy-ASE}
\end{figure}

\paragraph{Mode-locking}
Mode-locked MIR fiber lasers emitting picosecond and femtosecond pulses are of particular interest recently driven by the potential for multiple practical applications.
Recent demonstration of a Ho:ZBLAN laser mode-locked using nonlinear polarization evolution (NPE) resulted in pulses of 180~fs with a peak power of 37~kW~\cite{Antipov2016}.
Subsequent application of this system proved its utility as a pump source for nonlinear optical phenomena in the form of few-cycle pulse production~\cite{Woodward2017a}, and wide spanning supercontinuum~\cite{Alamgir2017}.
This system was preceded by Er NPE lasers that were negatively impacted by the influence of atmospheric absorption due to water vapour in the free space propagation segments of the cavity, contributing to longer duration pulses~\cite{Duval2015,Hu2015,Duval2016}.
While the slightly longer peak emission wavelength of Ho as compared to Er mitigates this issue to a degree, dysprosium would represent continued improvement on this front as the emission spectrum is yet further removed from the atmospheric water absorption peak in the region of 2.7-2.8~\si{\micro\m}.
Further, as the limit of attainable pulse duration is to first order inversely proportional to the gain bandwidth, dysprosium offers additional potential advantage due to a comparatively large emission spectrum.
Directly measured amplified spontaneous emission (ASE) spectra of Ho and Dy:ZBLAN fibers shown in Figure~\ref{fig-HoDy-ASE} illustrate this striking difference.

While pulses of picosecond duration have been generated in Er and Ho:ZBLAN systems employing a wide variety of saturable absorber materials~\cite{Zhu2017a}, thus far the saturable absorber results in Dy:ZBLAN have been limited to the Q-switched operation outlined above.
A novel alternative approach to successful picosecond pulse generation in MIR Dy fiber lasers is frequency shifted feedback~\cite{Woodward2018}.
While complete description of the mechanism can be found elsewhere (see e.g. \cite{Sabert1994}), in brief, with a frequency shifting element placed in the cavity, light is continually shifted monotonically each round trip, eventually shifting outside the gain bandwidth and inducing loss.
With sufficient nonlinearity (as is frequently encountered in a fiber geometry), intense light undergoing a frequency shift broadens by self phase modulation (SPM), spreading some of its energy back towards the center of the gain bandwidth. 
This serves as an intensity discriminating process, allowing for the buildup of short intense pulses as they experience higher net gain than low intensity CW light.
Additionally, as the light generated by SPM is phase coherent, and the frequency shift passes the phase of each spectral component to its shifted neighbor, the pulse produced is coherent, resembling the classic features of conventional mode-locking.
Thus FSF lasers are generally labelled as `mode-locked' systems despite strictly being a misnomer.

\begin{figure}[!htbp]
	\centering
	\begin{overpic}{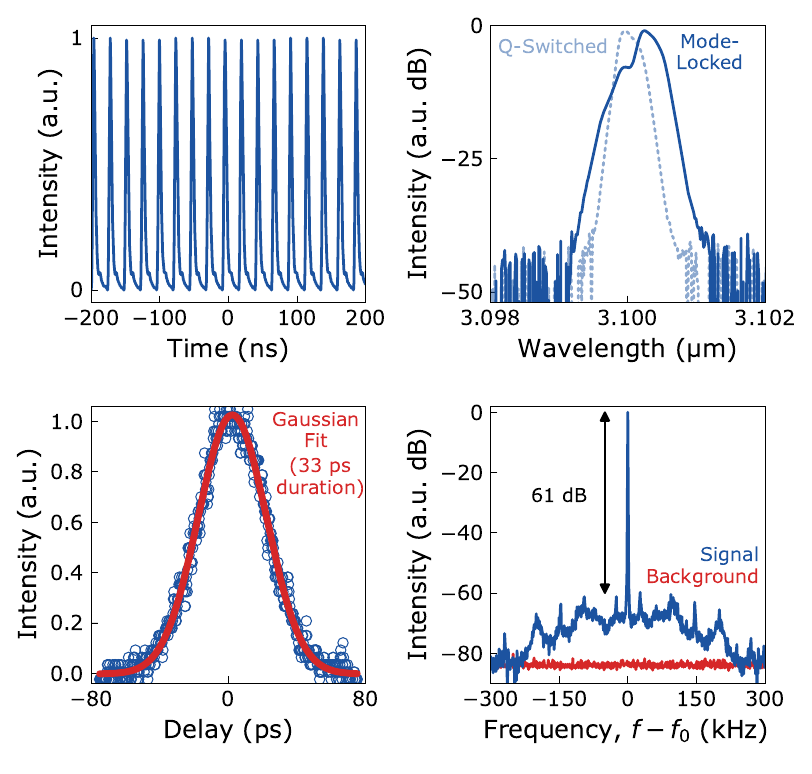}
	\put(1,91.5){(a)}
	\put(47,91.5){(b)}
	\put(1,46){(c)}
	\put(47,46){(d)}
	\end{overpic}
	\caption{Experimental results of the FSF Dy:ZBLAN fiber laser: (a) mode-locked pulse train; (b) optical spectrum; (c) autocorrelation trace; (d) RF spectrum, relative to fundamental repetition rate $f_0=44.5$~MHz. Data taken from Ref.~\cite{Woodward2018}.}
	\label{fig-FSF}
\end{figure}
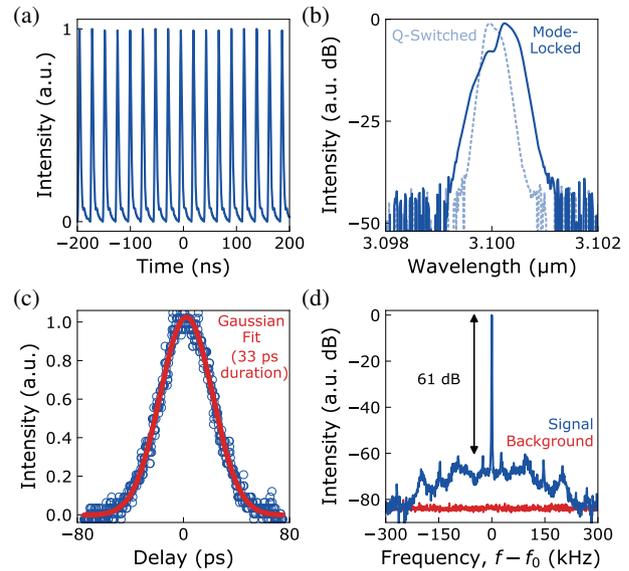

The FSF Dy:ZBLAN cavity arrangement is identical in components to the swept-wavelength system illustrated in Figure~\ref{fig-AOTF-NH3}(a) as the AOTF imparts a frequency shift equal to the RF drive frequency generating the acoustic wave in the crystal.
Experimental results from the demonstration of a picosecond `mode-locked' FSF Dy:ZBLAN fiber laser are seen in Figure~\ref{fig-FSF}.
It should be noted that in a FSF cavity pure CW operation is essentially inhibited as the frequency shift disrupts the establishment of longitudinal modes, in this case resulting in self-Q-switched operation at pump powers below 500~mW.
At higher pump powers, in the `mode-locked' regime, with the AOTF drive frequency set at 18.1~MHz (equating to a center diffraction wavelength of 3.1~\si{\micro\m}) a stable train of pulses is produced at the cavity round trip time (see Figure~\ref{fig-FSF}(a)).
Picosecond pulsed operation is confirmed by observation of spectral broadening, high signal to noise contrast in the fundamental RF spectrum of photodiode signal, and conclusively with an intensity autocorrelation revealing a pulse of 33~ps duration  (see Figure~\ref{fig-FSF}(b-d)). 

As described, the FSF mechanism requires a spectral filtering effect, which can either be the gain bandwidth itself, or an explicitly inserted filtering cavity element.
In the demonstrated case the AOTF serves as the limiting filter, with a FWHM passband of nominally 5~nm, significantly narrower than the dysprosium gain.
For this reason pulse durations from the FSF Dy:ZBLAN setup are lower limited in duration, and despite anomalous dispersion in the ZBLAN fiber in this wavelength range, soliton shaping is not found to influence the pulse properties~\cite{Woodward2018}.
Recently a similar demonstration of FSF-pulsed operation in Ho:ZBLAN utilized an AOM in place of a narrowband AOTF.
Here the wider effective filter bandwidth enabled an order of magnitude reduction in pulse duration, thus FSF in Dy systems have prospects for ultrafast pulse generation~\cite{DeSterke1995,Majewski2019}.

\begin{figure}[!htbp]
	\includegraphics{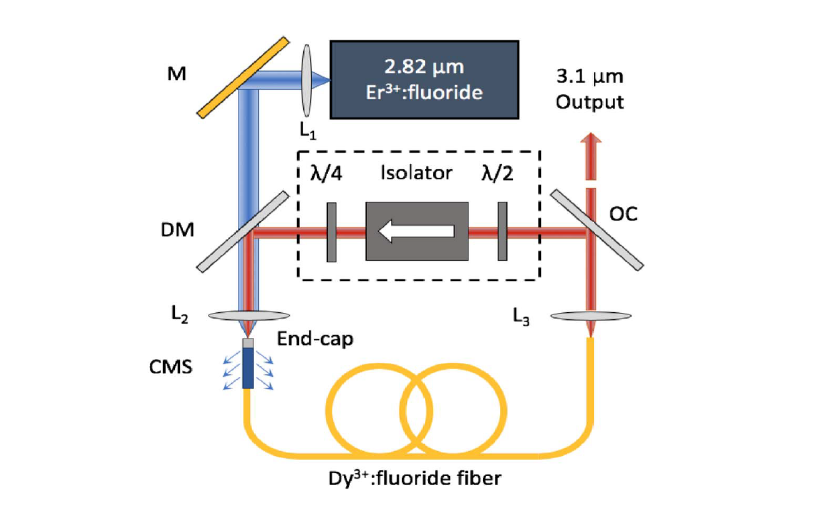}
	\caption{Experimental layout of a Dy NPE mode-locked laser. Reproduced with permission~\cite{Wang2019}. 2019, OSA.}
	\label{fig-NPE}
\end{figure}

\begin{figure}[!htbp]
	\begin{overpic}{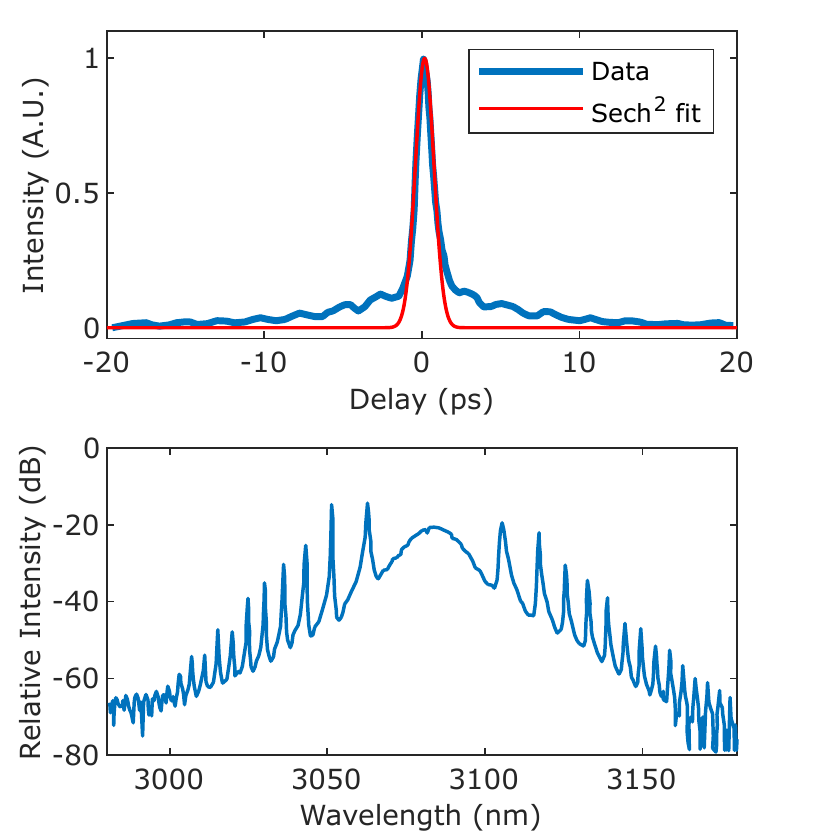}
		\put(2,95){(a)}
		\put(2,50){(b)}
		
	\end{overpic}
	\caption{Experimental results from a Dy:ZBLAN fiber laser mode-locked using nonlinear polarization evolution: (a) intensity autocorrelation trace with hyperbolic secant fit; (b) optical spectrum of the mode-locked pulse, exhibiting strong Kelly sidebands corresponding to dispersive waves. Data taken from Ref.~\cite{Wang2019}.}
	\label{fig-NPE-data}
\end{figure}

For operation in the solitonic regime, shorter pulses are required and are successfully generated using the NPE technique implemented in the cavity seen in Figure~\ref{fig-NPE}~\cite{Wang2019}.
This system employs the efficient in-band pumping scheme, and though requires a free space segment of the cavity is largely immune from the influence of atmospheric water absorption due the lasing wavelength of 3.1~\si{\micro\m} as defined by the coatings applied to the input dichroic and output coupler mirrors.
Pulse durations from this system (see Figure~\ref{fig-NPE-data}(a)) overcome the sub-1~ps barrier for the first time, reaching as short as 828~fs with pulse energies up to 4.8~nJ.
A measured optical spectrum (see Figure~\ref{fig-NPE-data}(b)) of the output indicate a spectral FWHM of 13.7~nm, representing only a fraction of the dysprosium gain bandwidth.
The authors note the presence of strong Kelly sidebands in the optical spectrum indicative of dispersive waves in the cavity and attribute the observable pedestal in the autocorrelation to this phenomenon.
They then suggest that compensation reducing the net dispersion would allow for the possibility of achieving substantially shorter pulses utilizing a greater portion fo the available gain.

\subsubsection{\SI{4}{\micro\m} emission}
\label{4micron_emission}
\begin{figure}[!htbp]
	\includegraphics{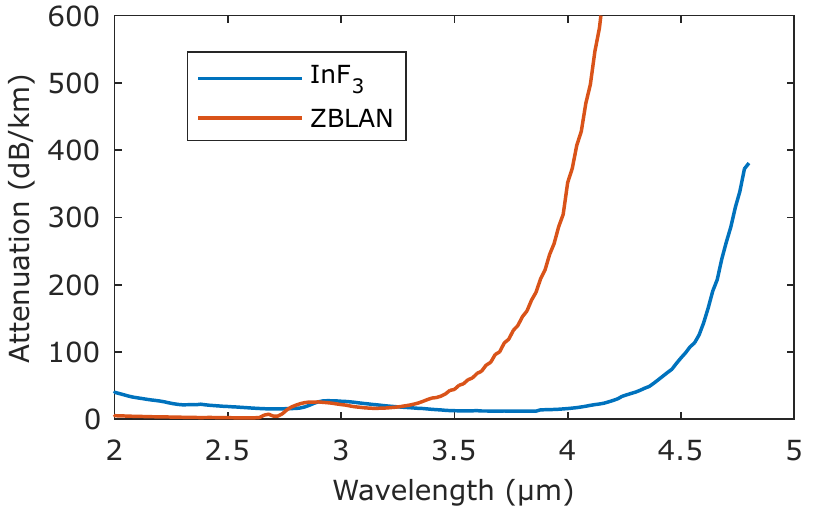}
	\caption{Measured attenuation of passive fluoride fibers. Data taken from Ref.~\cite{Majewski2018b}.}
	\label{fig-InF3}
\end{figure}

All of the fiber laser results thus far presented have operated on the ${^6H_{13/2}\rightarrow^6H_{15/2}}$ transition around 3~\si{\micro\m}, specifically utilizing ZBLAN as the host medium.
As we have outlined, the  4~\si{\micro\m} ${^6H_{11/2}\rightarrow^6H_{13/2}}$ transition also offers the potential for high efficiency and in fact is the majority focus of bulk solid state architecture  research effort.
However ZBLAN is not a suitable host at these wavelengths due to the high levels of multiphonon absorption.
Indium fluoride glasses (InF$_3$) on the other hand exhibit a longer multiphonon edge and increased transparency in the 4~\si{\micro\m} spectral range, as is evidenced by measured fiber attenuation data presented in Figure~\ref{fig-InF3}.

\begin{figure}[!htbp]
	\includegraphics{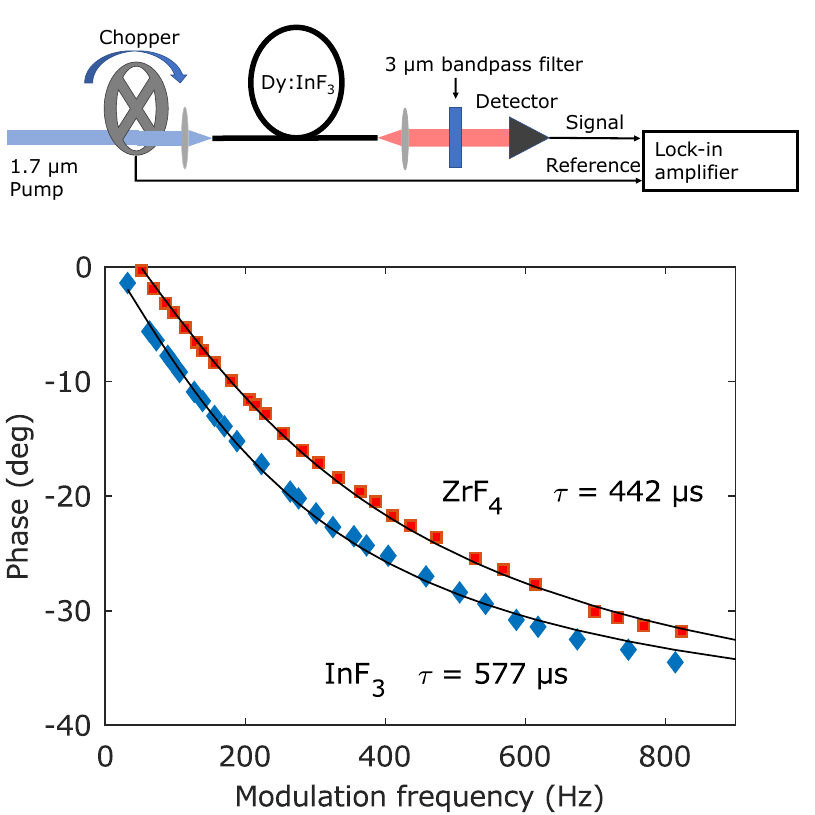}
	\caption{Frequency domain measurement of  $^6H_{13/2}$  fluorescence lifetime in both Dy:ZBLAN and Dy:InF$_3$ fibers. Data taken from Ref.~\cite{Majewski2018b}.}
	\label{fig-InF3_lifetime}
\end{figure}

Beyond the issue of fiber attenuation is that this transition is intrinsically self-terminating and particularly so in ZBLAN as non-radiative decay render $\tau_{upper}<<\tau_{lower}$.
Measurement of non-radiative rate ($W_{nr}$) from the $^6H_{11/2}$ level is challenging as the emission is weak and decays rapidly, requiring fast sensitive detection for a direct fluorescence decay determination.
Indirect measurement in bulk InF$_3$ glass using the integrated emission from the first three excited states back to ground indicate that $W_{nr}$ is actually slightly larger in InF$_3$~\cite{Quimby2014}.
Requirements placed on the detection system can be relaxed if measurement is done alternatively in the frequency domain, allowing for the possibility of  direct measurement of emission lifetime (and corollary $W_{nr}$) in a fiber geometry~\cite{Brunel1996}.
Here, instead of populating excited states rapidly with intense pump pulses, a CW pump beam is modulated and the phase of fluorescence response as a function of modulation frequency is represented by the relation  $\phi(\omega)=\arctan(\omega\tau)$.
A schematic of an experimental implementation of this method  is seen in Figure~\ref{fig-InF3_lifetime}(a). Unfortunately measurement of $^6H_{11/2}$ level required pump modulation in excess of that possible with the chosen method of mechanical chopping.
However a comparison measurement of $^6H_{13/2}$ lifetime in both ZBLAN and InF$_3$ reveal the expected increase in observed fluorescence lifetime due to the reduction of $W_{nr}$ (see Figure~\ref{fig-InF3_lifetime}(b))~\cite{Majewski2018b}.

\begin{figure}[!htbp]
	\includegraphics{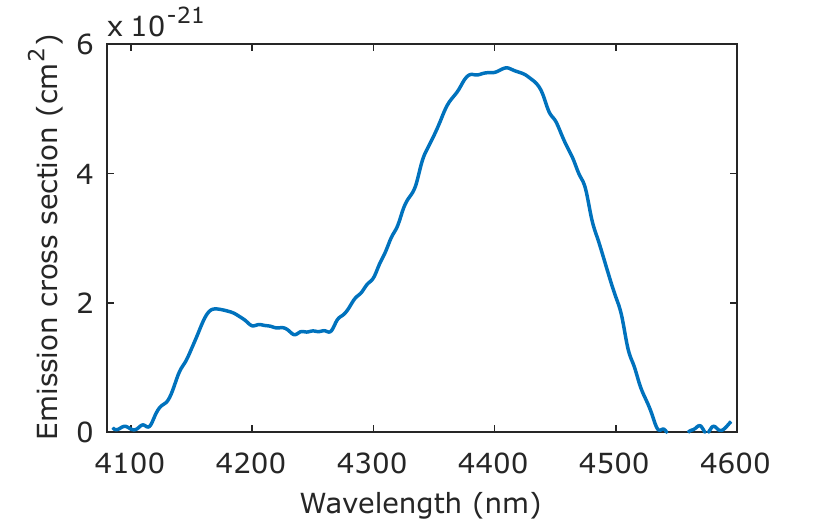}
	\caption{Emission spectrum of the ${^6H_{11/2}\rightarrow^6H_{13/2}}$ transition in Dy:InF$_3$ fiber. Data taken from Ref.~\cite{Majewski2018b}. }
	\label{fig-InF3_ASE}
\end{figure}

Despite a reduction of non-radiative rate in InF$_3$ as compared to ZBLAN, the upper state 4~\si{\micro\m} lifetime remains more than an order of magnitude smaller than that of the lower state, placing heavy constraint on the possibility of CW lasing.
One proposed solution is a cascade lasing scheme, using 3~\si{\micro\m} stimulated emission to de-populate $^6H_{13/2}$ and reduce the effective lifetime to a degree where CW lasing on the ${^6H_{11/2}\rightarrow^6H_{13/2}}$ transition is possible.
Numerical study of this approach in InF$_3$ shows that even if $W_{nr}$ is comparable in both glasses, the decrease in attenuation makes InF$_3$ much more promising~\cite{Quimby2013}.
However, this approach requires highly customized cavity optics or FBGs and even under optimal conditions a substantial amount of pump power to reach oscillation threshold.
Thus in practice lasing is only achieved on the 3~\si{\micro\m} transition, though ${^6H_{13/2}\rightarrow^6H_{15/2}}$ fluorescence is directly observed, representing the first emission beyond a longstanding 4~\si{\micro\m} barrier in fluoride glass fibers~\cite{Majewski2018b}.
The fluorescence spectrum acquired is presented in Figure~\ref{fig-InF3_ASE}, where the long wavelength emission edge is substantially sharper than that seen for example in a sulfide glass (see e.g. Figure~\ref{fig-glass-emission}(b)) due to the onset of multiphonon absorption.
To properly implement the F\"{u}chtbauer-Ladenburg relation and extract numerical values for emission cross section, one would need to accurately account for this effect and the obvious contribution from CO$_2$ absorption around 4.2~\si{\micro\m}.

Theoretically, a more compelling solution to achieving fiber laser emission from the ${^6H_{11/2}\rightarrow^6H_{13/2}}$ transition is the use of chalcogenide glass as host.
Comparing calculated and measured fluorescence lifetimes in various chalcogenide glasses (see Table.~\ref{lifetime_table_glass}) it is readily apparent that the reduction due to non-radiative decay is dramatically minimized as compared to ZBLAN.
While the 4~\si{\micro\m} transition remains intrinsically self-terminating in nature, a chalcogenide host renders this condition increasingly surmountable.
Indeed several authors have considered numerically a cascade lasing scheme in Dy:chalcogenide glass fiber, pumping the upper laser level directly at 1.7~\si{\micro\m}~\cite{Quimby2008,Sujecki2010,Falconi2017,Guo2018}. 
It is found that the substantial increase in upper state lifetime, coupled with the reduction in $\tau_{lower}/\tau_{upper}$ ratio results in much greater design flexibility in terms of cavity mirror requirements, i.e. much larger outcoupling ratios are possible.
Additionally, values of pump power at threshold reduce to much more technically feasible values less than 1~W.

However, thus far in practice doped chalcogenide fiber has proven unsuitable as a medium for coherent laser emission. 
Largely this is due to inability to support high doping concentrations of rare earths, and a background loss value that is substantially higher than ZBLAN, even in current state of the art chalcogenide fibers.
While improvement in these areas is ongoing in the field of chalcogenide glass fabrication, Dy-doped chalcogenide fibers already find utility in gas detection demonstrations when used simply as an ASE fluorescence source~\cite{Starecki2015,Ari2018,Pele2016}.

	\section{Conclusion and Outlook}
	\label{sec:outlook}
	
In this review, we have examined the key spectroscopic parameters and important developments in the use of the Dy$^{3+}$ ion as the active material in MIR lasers. 	
The Dy$^{3+}$ energy level structure relevant to the MIR is characterized essentially by a cluster of closely spaced levels from 0~\si{\per\cm} to approximately 13,000~\si{\per\cm}, meaning the choice of host material and number of electronic transitions leading to laser emission is limited. 
This problem has stifled progress, particularly when compared to the other rare earths that provide a variety of electronic transitions that emit high efficiency laser radiation.
That said, the combination of a strong interest in the MIR, wider availability of convenient pump sources and the development of new low-loss and robust host materials has created a resurgence in interest in the dysprosium ion as an efficient source of MIR light. 
This is evidenced by the recent sharp increase in Dy$^{3+}$-doped laser publications (particularly those using fiber media) and numerous recent landmark demonstrations, which we now reflect upon to consider the role of dysprosium and its future prospects in the wider landscape of MIR sources.

In terms of spectroscopy, \SI{3}{\micro\m} (${^6H_{13/2}\rightarrow^6H_{15/2}}$) and \SI{4.3}{\micro\m} (${^6H_{11/2}\rightarrow^6H_{13/2}}$) emission has been thoroughly explored in a number of crystalline hosts and various excitation wavelengths have been identified.
PGS has shown to be particularly promising to date as a host material\cite{Doroshenko2009, Jan2010, Jelinkova2011, Jelinkova2013a, Jelinkova2016} although it is noteworthy that new low-phonon-energy hosts continue to be investigated (e.g. Lu$_2$O$_3$~\cite{Heuer2018}), potentially leading to even more favourable properties.
Co-doping (with Yb$^{3+}$, Er$^{3+}$ and Tm$^{3+}$~\cite{Wang2017,Wang2018,Dexter1953,Tarelho1997,Shen2018,Ye2017,Toncelli1999,Tigreat2001,Hu2019,Doualan2018,Li2011,Ma2015,Tian2012,Tian2012a,Heo1997}) is also an area to have received research attention, highlighting routes to enhanced \SI{3}{\micro\m} emission efficiency, although these ideas have unfortunately yet to translate into laser experiments.
This is therefore an area for further study, in addition to consideration of other co-doping schemes to enhance the longer wavelength Dy$^{3+}$ \SI{4}{\micro\m} transition.

Beyond crystals, researchers have considered Dy$^{3+}$-doped glasses.
Silica is precluded from consideration due to its high phonon energy, leading to interest in soft glasses, particularly fluorides.
ZBLAN has been the most thoroughly investigated glass here, although this has had the unfortunate consequence of actually limiting study to the \SI{3}{\micro\m} transition due to ZBLAN's transparency cut-off at around \SI{3.8}{\micro\m}.
Recent developments using lower phonon energy InF$_3$ glasses are now changing this, however: preliminary work has already observed fluorescence beyond \SI{4}{\micro\m} from Dy$^{3+}$:InF$_3$~\cite{Majewski2018b} but significant work remains e.g. to quantify the $^6H_{11/2}$ lifetime and identify solutions to overcome the self-terminating nature of the \SI{4}{\micro\m} transition.
Many spectroscopic studies on tellurite and chalcogenide glasses have also been reported, although these have not yet led to laser demonstrations, even at \SI{3}{\micro\m}, which could be attributed to their higher loss and inability to achieve high doping concentrations with current generation materials / manufacturing.
The improvement of doped glass and fiber fabrication techniques is therefore also an important ongoing research topic.

In terms of dysprosium laser technology, despite early interest in \SI{3}{\micro\m} Dy$^{3+}$-doped crystalline laser sources, the majority of crystalline laser work has focussed on \SI{4.3}{\micro\m} emission. 
The challenging spectroscopy (i.e self-terminating transition) has resulted in almost all studies using pulsed pump schemes, although impressive mJ-level outputs have been reported. 
It is notable, however, that CW lasing at \SI{4.3}{\micro\m} has been achieved by exploiting pump ESA from the lower laser level to overcome the self-terminating nature of the transition~\cite{Jelinkova2013a}---this novel idea demonstrates that with further spectroscopic study to better understand ESA effects, higher performance power-scalable dysprosium crystalline lasers are to be expected.

It is Dy$^{3+}$-doped \emph{fibers}, however, where technological progress has been most rapid, albeit limited exclusively to the \SI{3}{\micro\m} transition.
Notably, in the last few years numerous Dy:ZBLAN fiber lasers performance landmarks have been reported, including: picosecond pulse generation beyond \SI{3}{\micro\m}~\cite{Woodward2018,Wang2019}, 10-W output power~\cite{Fortin2019}, 73\% slope efficiency~\cite{Woodward2018b} and $>$550-nm tunability~\cite{Majewski2018a}.
These demonstrations also surpass many performance records of the more established Er$^{3+}$ and Ho$^{3+}$ MIR fiber laser ions.
By comparison, Dy$^{3+}$ offers a number of intrinsic advances, including significantly broader emission cross-section and the option for in-band pumping with theoretical Stokes efficiency exceeding 90\% (\SI{3}{\micro\m} emission from Er$^{3+}$ and Ho$^{3+}$ is from higher lying levels in their energy level structure, requiring shorter wavelength pump sources and thus yielding Stokes-limited efficiencies of only 35\% and 40\%, respectively).
The requirement for a MIR pump source for in-band-pumped Dy$^{3+}$ lasers is a complication, compared to simpler diode pump schemes for other ions, although it is noted that Er$^{3+}$:ZBLAN fiber lasers are a rapidly maturing technology which are ideal for this purpose, since they can be spliced directly to Dy$^{3+}$:ZBLAN fibers to form monolithic robust systems~\cite{Fortin2019}.
While such Er$^{3+}$ lasers are currently limited in power to $\sim$40~W by thermomechanical failure effects~\cite{Aydn2018}, Dy$^{3+}$ offers an opportunity to circumvent this limit through the higher system efficiency (thus, lower thermal load).
For example, taking advantage of the broad Dy$^{3+}$ absorption cross section, numerous Er$^{3+}$:ZBLAN fiber lasers could be spectrally multiplexed to pump a dysprosium laser, resulting in an extremely high brightness source, possibly exceeding 100~W.

While bulk lasers have generated pulses with mJ-level energies and long ($>$\SI{}{\micro\s}) durations, fiber lasers are better suited for short-pulse operation.
For example, mode-locked Dy$^{3+}$:ZBLAN lasers have been reported with pulse duration down to 828~fs (nJ-level energy)~\cite{Wang2019}.
Nonetheless, given the broad Dy$^{3+}$ gain bandwidth (which ultimately limits achievable pulse durations via the Fourier transform limit), there is scope for $<$100~fs few-cycle pulse generation directly from mode-locked dysprosium cavities, providing dispersion is carefully managed.
The longer wavelength compared to other MIR fiber laser ions is also advantageous, avoiding atmospheric water vapour absorption which is particularly problematic for ultrafast Er$^{3+}$ lasers.
However, the potential of Dy$^{3+}$ for high energy Q-switched laser development is comparatively worse than Er$^{3+}$ and Ho$^{3+}$, due to the shorter upper state lifetime and thus, reduced gain storage.

Alongside reports of experimental dysprosium lasers, numerous studies have addressed the problem of optimising such systems and accessing new parameter regimes numerically.
This has led, for example, to proposed designs for broader tunability~\cite{Woodward2018a}, augmented pulse energies~\cite{Woodward2019,Falconi2017} and \SI{4}{\micro\m} emission by cascade lasing~\cite{Quimby2013}.
There have also been a number of works simulating Dy$^{3+}$:chalcogenide lasers~\cite{Sujecki2010,Quimby2008,Guo2018} based on spectroscopic measurements.
These works offer useful guidelines for future experiments, although further work is obviously required to verify these numerical predictions, possibly including amendments to simulations to account for currently undiscovered phenomena.

The emergence of dysprosium lasers will also open new optical applications.
For example, the broad emission cross section of the \SI{3}{\micro\m} transition has already been exploited to develop a swept-wavelength laser for real-time sensing~\cite{Woodward2018a}, since Dy$^{3+}$'s emission range spans a number of important functional groups (e.g. OH, NH, CH moieties).
Further sensing technologies could be expected from operation at longer wavelength, e.g. near CO$_2$ absorption at \SI{4.2}{\micro\m}.
In addition, the development of high-power CW sources near \SI{3.4}{\micro\m} can target resonant absorption of CH bonds, enabling cutting and micro-machining of technical materials such as polymers.
For practical realization of such applications (i.e. deployment outside the laboratory), dysprosium lasers need to be compact, robust and reliable.
Fortunately, with the strong community interest in the MIR, complementary optical components and techniques such as FBGs~\cite{Bernier2007,Bharathan2017}, couplers~\cite{Stevens2016,Tavakoli2017} and low-loss splicing~\cite{Thapa2015} are also under development, paving the way to complete turn-key MIR systems.

The future prospects for dysprosium are therefore very strong. 
In the coming years, with improved low-loss host material fabrication, we expect to see Dy-doped laser slope efficiencies exceeding 90\% around \SI{3}{\micro\m} and further tunability extension to exploit the full 2.5 to \SI{3.5}{\micro\m} emission cross section.
Exploration of novel host materials will also offer practical access to the longer wavelength transitions.
With multiplexed in-band pumping, Dy$^{3+}$ is the most promising rare-earth candidate for developing 100-W class MIR fiber lasers.
Such tunable and high-power sources are already ideal for remote sensing and polymer processing applications, which are expected to be further explored and exploited, extending the proof-of-principle application demonstrations to date.
Finally, dispersion-managed mode-locked Dy$^{3+}$ cavity designs will enable high-power sub-100 fs pulse generation, ideally suited for pumping nonlinear chalcogenide fiber for supercontinuum generation and other nonlinear optical applications.

To conclude, dysprosium is now firmly established as an important ion for MIR laser development. 
This review has highlighted the broad parameter space spanned by dysprosium-doped lasers to date, and also the potential for further developments owing to the ion's unique spectroscopy.
Dysprosium is certainly set to play a valuable role in opening up the mid-infrared region and the creation of important new applications across science, medicine and manufacturing.

\begin{acknowledgement}
This work has been supported by funding from the Asian Office of Aerospace Research and Development, grant FA2386-19-1-0043	
\end{acknowledgement}

\begin{biographies}
	\authorbox{cv_Majewski}{Matthew R. Majewski}{received a MS from Northeastern University in 2012. Prior to joining Macquarie University in 2014 he was member of the high power laser development team at Silex Systems (Australia). He completed his PhD thesis on dysprosium mid-infrared fiber lasers in 2017 and is currently a research fellow with the Mid-infrared Fiber Sources Group. His current research focus is the development of novel high power and ultrafast fiber laser sources.}
	
	\authorbox{cv_Woodward}{Robert I. Woodward}{received an MEng from Trinity Hall, University of Cambridge (UK) in 2012, followed by a PhD in laser physics and nonlinear fiber optics at Imperial College London (UK).
	 He moved to Macquarie University (Australia) in 2016 to take up an MQ Research Fellowship to continue his fiber laser research, with a particular focus on the MIR region.}
	
	\authorbox{cv_Jackson}{Stuart D. Jackson}{received his PhD from Macquarie University in 1990.
		 After postdoctoral work at the University of Manchester (UK) he joined the University of Sydney where he became a Senior Research Fellow and subsequently the leader of the mid-infrared photonics project within the Centre for Ultrahigh-bandwidth Devices for Optical Systems (CUDOS). He is currently a professor at Macquarie University where he serves as director of the Mid-infrared Fiber Sources Group and deputy director of the MQ Photonics Research Center. He is a member of the OSA and a senior member of the IEEE.}
\end{biographies}

\providecommand{\WileyBibTextsc}{}
\let\textsc\WileyBibTextsc
\providecommand{\othercit}{}
\providecommand{\jr}[1]{#1}
\providecommand{\etal}{~et~al.}


\begin{thebibliography}{[100]}
	
	\bibitem{Dieke1961}
	\textsc{G.\,H. Dieke},  \textsc{H.\,M. Crosswhite},  and  \textsc{B.~Dunn},
	\jr{J. Opt. Soc. Am.} \textbf{51}(8), 820 (1961).
	
	
	\bibitem{Kaminskii2002}
	\textsc{D.~Temple},  \textsc{A.\,A. Kornienko},  \textsc{A.\,A. Kaminskii},
	\textsc{J.\,B. Gruber},  \textsc{K.\,i. Ueda},  \textsc{R.\,F. Klevtsova},
	\textsc{U.~H{\"{o}}mmerich},  \textsc{A.\,A. Pavlyuk},  \textsc{S.\,N.
		Bagaev},  \textsc{F.\,A. Kuznetsov},  \textsc{E.\,B. Dunina},
	\textsc{B.~Zandi},  and  \textsc{J.\,T. Seo},
	\jr{Phys. Rev. B} \textbf{65}(12), 125108 (2002).
	
	
	\bibitem{Parisi2005}
	\textsc{D.~Parisi},  \textsc{A.~Toncelli},  \textsc{M.~Tonelli},
	\textsc{E.~Cavalli},  \textsc{E.~Bovero},  and  \textsc{A.~Belletti},
	\jr{J. Phys. Condens. Matter} \textbf{17}(17), 2783--2790 (2005).
	
	
	\bibitem{Macalik1998}
	\textsc{L.~Macalik},  \textsc{J.~Hanuza},  \textsc{B.~Macalik},
	\textsc{W.~Ryba-Romanowski},  \textsc{S.~Golab},  and
	\textsc{A.~Pietraszko},
	\jr{J. Lumin.} \textbf{79}(1), 9--19 (1998).
	
	
	\bibitem{Gruber1998}
	\textsc{J.\,B. Gruber},  \textsc{B.~Zandi},  and  \textsc{L.~Merkle},
	\jr{J. Appl. Phys.} \textbf{83}(2), 1009--1017 (1998).
	
	
	\bibitem{Limpert2000}
	\textsc{J.~Limpert},  \textsc{H.~Zellmer},  \textsc{P.~Riedel},
	\textsc{G.~Maze},  and  \textsc{A.~Tunnermann},
	\jr{Electron. Lett.} \textbf{36}(16), 1386--1387 (2000).
	
	
	\bibitem{Fujimoto2010}
	\textsc{Y.~Fujimoto},  \textsc{O.~Ishii},  and  \textsc{M.~Yamazaki},
	\jr{Electron. Lett.} \textbf{46}(8), 586 (2010).
	
	
	\bibitem{Bowman2012}
	\textsc{S.\,R. Bowman},  \textsc{S.~O'Connor},  and  \textsc{N.\,J. Condon},
	\jr{Opt. Express} \textbf{20}(12), 12906--11 (2012).
	
	
	\bibitem{Bolognesi2014}
	\textsc{G.~Bolognesi},  \textsc{D.~Parisi},  \textsc{D.~Calonico},
	\textsc{G.\,A. Costanzo},  \textsc{F.~Levi},  \textsc{P.\,W. Metz},
	\textsc{C.~Kr{\"{a}}nkel},  \textsc{G.~Huber},  and  \textsc{M.~Tonelli},
	\jr{Opt. Lett.} \textbf{39}(23), 6628 (2014).
	
	
	\bibitem{Sardar2004}
	\textsc{D.\,K. Sardar},  \textsc{W.\,M. Bradley},  \textsc{R.\,M. Yow},
	\textsc{J.\,B. Gruber},  and  \textsc{B.~Zandi},
	\jr{J. Lumin.} \textbf{106}(3-4), 195--203 (2004).
	
	
	\bibitem{Seltzer1996}
	\textsc{M.\,D. Seltzer},  \textsc{A.\,O. Wright},  \textsc{C.\,A. Morrison},
	\textsc{D.\,E. Wortman},  \textsc{J.\,B. Gruber},  and  \textsc{E.\,D.
		Filer},
	\jr{J. Phys. Chem. Solids} \textbf{57}(9), 1175--1182 (1996).
	
	
	\bibitem{Page1997}
	\textsc{R.\,H. Page},  \textsc{K.\,I. Schaffers},  \textsc{S.\,A. Payne},  and
	\textsc{W.\,F. Krupke},
	\jr{J. Light. Technol.} \textbf{15}(5), 786--793 (1997).
	
	
	\bibitem{Tkachuk1999a}
	\textsc{A.\,M. Tkachuk},  \textsc{S.\,V. Ivanova},  \textsc{L.\,I. Isaenko},
	\textsc{A.\,P. Yelisseyev},  \textsc{S.\,A. Payne},  \textsc{R.\,W. Solarz},
	\textsc{M.\,C. Nostrand},  and  \textsc{R.\,H. Page},
	\jr{Acta Phys. Pol. A} \textbf{95}(3), 381--394 (1999).
	
	
	\bibitem{Tanabe1995}
	\textsc{S.~Tanabe},  \textsc{T.~Hanada},  \textsc{M.~Watanabe},
	\textsc{T.~Hayashi},  and  \textsc{N.~Soga},
	\jr{J. Am. Ceram. Soc.} \textbf{78}(11), 2917--2922 (1995).
	
	
	\bibitem{Ballato1997}
	\textsc{J.~Ballato},  \textsc{R.\,E. Riman},  and  \textsc{E.~Snitzer},
	\jr{Opt. Lett.} \textbf{22}(10), 691 (1997).
	
	
	\bibitem{Amplifiers1998}
	\textsc{D.\,T. Schaafsma},  \textsc{L.\,B. Shaw},  \textsc{B.~Cole},
	\textsc{J.\,S. Sanghera},  and  \textsc{I.\,D. Aggarwal},
	\jr{IEEE Photonics J.} \textbf{10}(11), 1548--1550 (1998).
	
	
	\bibitem{Li2017}
	\textsc{S.~Li},  \textsc{L.~Zhang},  \textsc{P.~Zhang},  \textsc{J.~Hong},
	\textsc{M.~Xu},  \textsc{T.~Yan},  \textsc{N.~Ye},  and  \textsc{Y.~Hang},
	\jr{Infrared Phys. Technol.} \textbf{87}, 65--71 (2017).
	
	
	\bibitem{Piramidowicz2008b}
	\textsc{R.~Piramidowicz},  \textsc{M.~Klimczak},  and  \textsc{M.~Malinowski},
	\jr{Opt. Mater. (Amst).} \textbf{30}(5), 707--710 (2008).
	
	
	\bibitem{Shestakova2007}
	\textsc{A.\,G. Okhrimchuk},  \textsc{L.\,N. Butvina},  \textsc{E.\,M. Dianov},
	\textsc{I.\,A. Shestakova},  \textsc{N.\,V. Lichkova},  \textsc{V.\,N.
		Zagorodnev},  and  \textsc{A.\,V. Shestakov},
	\jr{J. Opt. Soc. Am. B} \textbf{24}(10), 2690 (2007).
	
	
	\bibitem{Jelinkova2016}
	\textsc{H.~Jelinkova},  \textsc{M.\,E. Doroshenko},  \textsc{V.\,V. Osiko},
	\textsc{M.~Jel{\'{i}}nek},  \textsc{J.~{\v{S}}ulc},  \textsc{M.~N{\v{e}}mec},
	\textsc{D.~Vyhl{\'{i}}dal},  \textsc{V.\,V. Badikov},  and  \textsc{D.\,V.
		Badikov},
	\jr{Appl. Phys. A Mater. Sci. Process.} \textbf{122}(8), 738 (2016).
	
	
	\bibitem{Johnson1973b}
	\textsc{L.\,F. Johnson} and  \textsc{H.\,J. Guggenheim},
	\jr{Appl. Phys. Lett.} \textbf{23}(2), 96--98 (1973).
	
	
	\bibitem{Davydova1977a}
	\textsc{M.\,P. Davydova},  \textsc{S.\,B. Zdanovich},  \textsc{B.\,N.
		Kazakov},  \textsc{S.\,L. Korableva},  and  \textsc{A.\,L. Stolov},
	\jr{Opt. Spectrosc} \textbf{42}(3), 327 (1977).
	
	
	\othercit
	\bibitem{Carnall}
	\textsc{W.\,T. Carnall} and  \textsc{H.\,M. Crosswhite},
	{Energy level structure and transition probabilities in the spectra of the
		trivalent lanthanides in LaF3},
	Tech. rep., Argonne National Lab, 1978.
	
	
	\bibitem{Grunberg1969}
	\textsc{P.~Grunberg},  \textsc{S.~Hufner},  \textsc{E.~Orlich},  and
	\textsc{J.~Schmitt},
	\jr{Phys. Rev.} \textbf{184}(2) (1969).
	
	
	\bibitem{Carnall1968}
	\textsc{W.\,T. Carnall},  \textsc{P.\,R. Fields},  and  \textsc{K.~Rajnak},
	\jr{J. Chem. Phys.} \textbf{49}(10), 4412--4423 (1968).
	
	
	\bibitem{Neogy1988}
	\textsc{D.~Neogy} and  \textsc{T.~Purohit},
	\jr{Phys. Status Solidi} \textbf{146}(1), 181--187 (1988).
	
	
	\othercit
	\bibitem{wortman1976rare}
	\textsc{D.\,E. Wortman},  \textsc{N.~Karayianis},  and  \textsc{C.\,A.
		Morrison},
	{Rare Earth Ion-Host Lattice Interactions. 6. Lanthanides in LiYF4.},
	Tech. rep., HARRY DIAMOND LABS ADELPHI MD, 1976.
	
	
	\bibitem{Hehlen2013}
	\textsc{M.\,P. Hehlen},  \textsc{M.\,G. Brik},  and  \textsc{K.\,W.
		Kr{\"{a}}mer},
	\jr{J. Lumin.} \textbf{136}, 221--239 (2013).
	
	
	\othercit
	\bibitem{Walsh2006}
	\textsc{B.\,M. Walsh},
	{Judd-Ofelt theory: principles and practices},
	in: Adv. Spectrosc. lasers Sens.,  (Springer, Dordrecht, 2006),
	pp.\,403--433.
	
	
	\bibitem{Xu1984}
	\textsc{L.\,W. Xu},  \textsc{H.\,M. Crosswhite},  and  \textsc{J.\,P.
		Hessler},
	\jr{J. Chem. Phys.} \textbf{81}(2), 698--703 (1984).
	
	
	\bibitem{Brik2004}
	\textsc{M.\,G. Brik},  \textsc{T.~Ishii},  \textsc{A.\,M. Tkachuk},
	\textsc{S.\,E. Ivanova},  and  \textsc{I.\,K. Razumova},
	\jr{J. Alloys Compd.} \textbf{374}(1-2), 63--68 (2004).
	
	
	\bibitem{Ivanova1999}
	\textsc{S.\,E. Ivanova} and  \textsc{A.\,M. Tkachuk},
	\jr{Opt. Spectrosc.} \textbf{87}(1), 42 (1999).
	
	
	\bibitem{Heuer2018}
	\textsc{A.\,M. Heuer},  \textsc{P.~von Brunn},  \textsc{G.~Huber},  and
	\textsc{C.~Krankel},
	\jr{Opt. Mater. Express} \textbf{8}(11) (2018).
	
	
	\bibitem{Wang2014}
	\textsc{Y.~Wang},  \textsc{J.~Li},  \textsc{Z.~Zhu},  \textsc{Z.~You},
	\textsc{J.~Xu},  and  \textsc{C.~Tu},
	\jr{Opt. Mater. Express} \textbf{4}(6), 1104 (2014).
	
	
	\bibitem{Nostrand1999}
	\textsc{M.\,C. Nostrand},  \textsc{R.\,H. Page},  \textsc{S.\,A. Payne},
	\textsc{W.\,F. Krupke},  and  \textsc{P.\,G. Schunemann},
	\jr{Opt. Lett.} \textbf{24}(17), 1215--1217 (1999).
	
	
	\bibitem{Nostrand2001}
	\textsc{M.\,C. Nostrand},  \textsc{R.\,H. Page},  \textsc{S.\,A. Payne},
	\textsc{L.\,I. Isaenko},  and  \textsc{A.\,P. Yelisseeyev},
	\jr{J. Opt. Soc. Am. B} \textbf{18}(3), 264--276 (2001).
	
	
	\bibitem{Mak1982}
	\textsc{A.\,A. Mak} and  \textsc{B.\,M. Antipenko},
	\jr{J. Appl. Spectrosc.} \textbf{37}(6), 1458--1471 (1982).
	
	
	\bibitem{Contact2012}
	\textsc{M.\,C. Nostrand},  \textsc{R.\,H. Page},  \textsc{S.\,A. Payne},
	\textsc{W.\,F. Krupke},  \textsc{P.\,G. Schunemann},  and  \textsc{L.\,I.
		Isaenko},
	\jr{OSA TOPS Adv. Solid State Lasers} \textbf{19}, 524--528 (1998).
	
	
	\bibitem{Hommerich2006}
	\textsc{U.~H{\"{o}}mmerich},  \textsc{E.~Nyein},  \textsc{J.\,A. Freeman},
	\textsc{P.~Amedzake},  \textsc{S.\,B. Trivedi},  and  \textsc{J.\,M. Zavada},
	\jr{J. Cryst. Growth} \textbf{287}(2), 230--233 (2006).
	
	
	\bibitem{Wetenkamp1992}
	\textsc{L.~Wetenkamp},  \textsc{G.\,F. West},  and  \textsc{H.~T{\"{o}}bben},
	\jr{J. Non. Cryst. Solids} \textbf{140}(C), 35--40 (1992).
	
	
	\bibitem{mcdougall1994judd-a}
	\textsc{J.~McDougall},  \textsc{D.~Hollis},  and  \textsc{M.~Payne},
	\jr{Phys. Chem. Glasses} \textbf{35}(6), 258--259 (1994).
	
	
	\bibitem{mcdougall1994judd}
	\textsc{J.~McDougall},  \textsc{D.~Hollis},  \textsc{X.~Liu},  and
	\textsc{M.~Payne},
	\jr{Phys. Chem. Glasses} \textbf{35}(3), 145--146 (1994).
	
	
	\bibitem{Adam1988}
	\textsc{J.\,L. Adam},  \textsc{A.\,D. Docq},  and  \textsc{J.~Lucas},
	\jr{J. Solid State Chem.} \textbf{75}(2), 403--412 (1988).
	
	
	\bibitem{orera1988optical}
	\textsc{V.~Orera},  \textsc{P.~Alonso},  \textsc{R.~Cases},  and
	\textsc{R.~Alcala},
	\jr{Phys. Chem. Glasses} \textbf{29}(2), 59--62 (1988).
	
	
	\bibitem{Cases1991}
	\textsc{R.~Cases} and  \textsc{M.\,A. Chamarro},
	\jr{J. Solid State Chem.} \textbf{90}, 313--319 (1991).
	
	
	\bibitem{Hormaldy1979}
	\textsc{J.~Hormaldy} and  \textsc{R.~Reisfeld},
	\jr{J. Non. Cryst. Solids} \textbf{30}, 337--348 (1979).
	
	
	\bibitem{Heo1996}
	\textsc{J.~Heo} and  \textsc{Y.\,B. Shin},
	\jr{J. Non. Cryst. Solids} \textbf{196}, 162--167 (1996).
	
	
	\bibitem{Wei1994}
	\textsc{K.~Wei},  \textsc{D.\,P. Machewirth},  \textsc{J.~Wenzel},
	\textsc{E.~Snitzer},  and  \textsc{G.\,H. Sigel},
	\jr{Opt. Lett.} \textbf{19}(12), 904--906 (1994).
	
	
	\bibitem{Sojka2017}
	\textsc{L.~Sojka},  \textsc{Z.~Tang},  \textsc{D.~Furniss},  \textsc{H.~Sakr},
	\textsc{Y.~Fang},  \textsc{E.~Beres-Pawlik},  \textsc{T.~Benson},
	\textsc{A.~Seddon},  and  \textsc{S.~Sujecki},
	\jr{J. Opt. Soc. Am. B Opt. Phys.} \textbf{34}(3), 70--79 (2017).
	
	
	\bibitem{Schweizer1996}
	\textsc{T.~Schweizer},  \textsc{D.\,W. Hewak},  \textsc{B.\,N. Samson},  and
	\textsc{D.\,N. Payne},
	\jr{Opt. Lett.} \textbf{21}(19), 1594--1596 (1996).
	
	
	\bibitem{Quimby2017}
	\textsc{R.\,S. Quimby},  \textsc{M.~Saad},  \textsc{R.\,S.\,Q. Uimby},
	\textsc{R.\,S. Quimby},  and  \textsc{M.~Saad},
	\jr{Opt. Lett.} \textbf{42}(1), 117--120 (2017).
	
	
	\bibitem{Richards2013}
	\textsc{B.\,D.\,O. Richards},  \textsc{T.~Teddy-Fernandez},  \textsc{G.~Jose},
	\textsc{D.~Binks},  and  \textsc{A.~Jha},
	\jr{Laser Phys. Lett.} \textbf{10}(8), 085802 (2013).
	
	
	\bibitem{Gomes2014}
	\textsc{L.~Gomes},  \textsc{J.~Lousteau},  \textsc{D.~Milanese},
	\textsc{E.~Mura},  and  \textsc{S.\,D. Jackson},
	\jr{J. Opt. Soc. Am. B} \textbf{31}(3), 429 (2014).
	
	
	\bibitem{Falconi2017}
	\textsc{M.\,C. Falconi},  \textsc{G.~Palma},  \textsc{F.~Starecki},
	\textsc{V.~Nazabal},  \textsc{J.~Troles},  \textsc{J.\,L. Adam},
	\textsc{S.~Taccheo},  \textsc{M.~Ferrari},  and  \textsc{F.~Prudenzano},
	\jr{J. Light. Technol.} \textbf{35}(2), 265--273 (2017).
	
	
	\bibitem{Gomes2010a}
	\textsc{L.~Gomes},  \textsc{A.\,F.\,H. Librantz},  and  \textsc{S.\,D.
		Jackson},
	\jr{J. Appl. Phys.} \textbf{107}(5) (2010).
	
	
	\bibitem{Shin1999}
	\textsc{Y.\,B. Shin} and  \textsc{J.~Heo},
	\jr{J. Non. Cryst. Solids} \textbf{253}(1-3), 23--29 (1999).
	
	
	\bibitem{Charpentier2013}
	\textsc{F.~Charpentier},  \textsc{F.~Starecki},  \textsc{J.\,L. Doualan},
	\textsc{P.~J{\'{o}}v{\'{a}}ri},  \textsc{P.~Camy},  \textsc{J.~Troles},
	\textsc{S.~Belin},  \textsc{B.~Bureau},  and  \textsc{V.~Nazabal},
	\jr{Mater. Lett.} \textbf{101}, 21--24 (2013).
	
	
	\bibitem{Falconi2016}
	\textsc{M.\,C. Falconi},  \textsc{G.~Palma},  \textsc{F.~Starecki},
	\textsc{V.~Nazabal},  \textsc{J.~Troles},  \textsc{S.~Taccheo},
	\textsc{M.~Ferrari},  and  \textsc{F.~Prudenzano},
	\jr{IEEE Photonics Technol. Lett.} \textbf{28}(18), 1984--1987 (2016).
	
	
	\bibitem{Shaw2001}
	\textsc{L.\,B. Shaw},  \textsc{B.~Cole},  \textsc{P.\,A. Thielen},
	\textsc{J.\,S. Sanghera},  and  \textsc{I.\,D. Aggarwal},
	\jr{IEEE J. Quantum Electron.} \textbf{37}(9), 1127--1137 (2001).
	
	
	\bibitem{Yang2017}
	\textsc{A.~Yang},  \textsc{J.~Qiu},  \textsc{M.~Zhang},  \textsc{H.~Ren},
	\textsc{C.~Zhai},  \textsc{S.~Qi},  \textsc{B.~Zhang},  \textsc{D.~Tang},
	and  \textsc{Z.~Yang},
	\jr{J. Alloys Compd.} \textbf{695}, 1237--1242 (2017).
	
	
	\bibitem{Nemec2000}
	\textsc{P.~N{\v{e}}mec},  \textsc{B.~Frumarov{\'{a}}},  \textsc{M.~Frumar},
	and  \textsc{J.~Oswald},
	\jr{J. Phys. Chem. Solids} \textbf{61}(10), 1583--1589 (2000).
	
	
	\bibitem{Zhou2016}
	\textsc{B.~Zhou},  \textsc{F.~Huang},  \textsc{M.~Cai},  \textsc{Y.~Tian},
	\textsc{J.~Zhou},  \textsc{S.~Xu},  and  \textsc{J.~Zhang},
	\jr{IEEE Photonics Technol. Lett.} \textbf{28}(4), 429--432 (2016).
	
	
	\bibitem{Li2012d}
	\textsc{J.~Li},  \textsc{L.~Gomes},  and  \textsc{S.\,D. Jackson},
	\jr{IEEE J. Quantum Electron.} \textbf{48}(5), 596--607 (2012).
	
	
	\bibitem{Sapir2016c}
	\textsc{O.\,H. Sapir},  \textsc{S.\,D. Jackson},  and  \textsc{D.~Ottaway},
	\jr{Opt. Lett.} \textbf{41}(7), 1676--1679 (2016).
	
	
	\bibitem{Iqbal1991}
	\textsc{T.~Iqbal},  \textsc{M.\,R. Shahriari},  \textsc{G.~Merberg},  and
	\textsc{G.\,H. Sigel},
	\jr{J. Mater. Res.} \textbf{6}(2), 401--406 (1991).
	
	
	\bibitem{Fujimoto2011}
	\textsc{Y.~Fujimoto},  \textsc{O.~Ishii},  and  \textsc{M.~Yamazaki},
	\jr{Solid State Lasers XX Technol. Devices} \textbf{7912}(February 2011),
	79120J (2011).
	
	
	\othercit
	\bibitem{Reisfeld1987}
	\textsc{R.~Reisfeld} and  \textsc{C.\,K. Jorgensen},
	{Excited State Phenomena in Vitreous Materials},
	in: Handb. Phys. Chem. Rare Earths,  (Elsevier, 1987),  pp.\,1--90.
	
	
	\bibitem{Lume1977}
	\textsc{C.\,B. Layne},  \textsc{W.\,H. Lowdermilk},  and  \textsc{M.\,J.
		Weber},
	\jr{Phys. Reivew B} \textbf{16}(1) (1977).
	
	
	\bibitem{Zhu2010b}
	\textsc{X.~Zhu} and  \textsc{N.~Peyghambarian},
	\jr{Adv. Optoelectron.} \textbf{2010} (2010).
	
	
	\bibitem{France1984}
	\textsc{P.~France},  \textsc{S.~Carter},  \textsc{J.~Williams},
	\textsc{K.~Beales},  and  \textsc{J.~Parker},
	\jr{Electron. Lett.} \textbf{20}(14), 607 (1984).
	
	
	\bibitem{Qi2017}
	\textsc{F.~Qi},  \textsc{F.~Huang},  \textsc{L.\,F. Zhou},  \textsc{Y.~Tian},
	\textsc{R.~Lei},  \textsc{G.\,Y. Ren},  \textsc{J.~Zhang},
	\textsc{L.~Zhang},  and  \textsc{S.~Xu},
	\jr{J. Lumin.} \textbf{190}(May), 392--396 (2017).
	
	
	\bibitem{Wang2017}
	\textsc{C.~Wang},  \textsc{Y.~Tian},  \textsc{H.~Li},  \textsc{Q.~Liu},
	\textsc{F.~Huang},  \textsc{B.~Li},  \textsc{J.~Zhang},  and  \textsc{S.~Xu},
	\jr{Infrared Phys. Technol.} \textbf{85}, 128--132 (2017).
	
	
	\bibitem{Wang2018}
	\textsc{T.~Wang},  \textsc{F.~Huang},  \textsc{G.~Ren},  \textsc{W.~Cao},
	\textsc{Y.~Tian},  \textsc{R.~Lei},  \textsc{J.~Zhang},  and  \textsc{S.~Xu},
	\jr{Opt. Mater. (Amst).} \textbf{75}, 875--879 (2018).
	
	
	\bibitem{Dexter1953}
	\textsc{D.\,L. Dexter},
	\jr{J. Chem. Phys.} \textbf{21}(5), 836--850 (1953).
	
	
	\bibitem{Tarelho1997}
	\textsc{L.~Tarelho},  \textsc{L.~Gomes},  and  \textsc{I.~Ranieri},
	\jr{Phys. Rev. B - Condens. Matter Mater. Phys.} \textbf{56}(22), 14344--14351
	(1997).
	
	
	\bibitem{Tigreat2001}
	\textsc{P.\,Y. Tigreat},  \textsc{J.\,L. Doualan},  \textsc{C.~Budasca},  and
	\textsc{R.~Moncorge},
	\jr{J. Lumin.} \textbf{95}, 23--27 (2001).
	
	
	\bibitem{Shen2018}
	\textsc{L.~Shen},  \textsc{N.~Wang},  \textsc{A.~Dou},  \textsc{Y.~Cai},
	\textsc{Y.~Tian},  \textsc{F.~Huang},  \textsc{S.~Xu},  and
	\textsc{J.~Zhang},
	\jr{Opt. Mater. (Amst).} \textbf{75}, 274--279 (2018).
	
	
	\bibitem{Ye2017}
	\textsc{R.~Ye},  \textsc{J.~Zhang},  \textsc{F.~Qi},  \textsc{Y.~Tian},
	\textsc{T.~Wang},  \textsc{S.~Xu},  \textsc{L.~Zhang},  \textsc{F.~Huang},
	and  \textsc{R.~Lei},
	\jr{Appl. Opt.} \textbf{56}(31), H24 (2017).
	
	
	\bibitem{Toncelli1999}
	\textsc{A.~Toncelli},  \textsc{M.~Tonelli},  \textsc{A.~Cassanho},  and
	\textsc{H.\,P. Jenssen},
	\jr{J. Lumin.} \textbf{82}(4), 291--298 (1999).
	
	
	\bibitem{Hu2019}
	\textsc{M.~Hu},  \textsc{Y.~Wang},  \textsc{Z.~Zhu},  \textsc{Z.~You},
	\textsc{J.~Li},  and  \textsc{C.~Tu},
	\jr{J. Lumin.} \textbf{207}(September 2018), 226--230 (2019).
	
	
	\othercit
	\bibitem{Doualan2018}
	\textsc{J.\,L. Doualan},  \textsc{A.~Benayad},  \textsc{A.~Braud},
	\textsc{P.~Camy},  and  \textsc{G.~Brasse},
	{Dy3+ doped CaF2 crystals spectroscopy for the development of Mid-infrared
		lasers around \SI{3}{\micro\m}},
	in: Fiber Lasers Glas. Photonics Mater. through Appl.,  (SPIE, 2018),
	p.\,1068329.
	
	
	\bibitem{Li2011}
	\textsc{Z.~Li},  \textsc{T.~Xu},  \textsc{X.~Shen},  \textsc{S.~Dai},
	\textsc{X.~Wang},  \textsc{Q.~Nie},  and  \textsc{X.~Zhang},
	\jr{J. Rare Earths} \textbf{29}(2), 105--108 (2011).
	
	
	\bibitem{Ma2015}
	\textsc{Y.~Ma},  \textsc{X.~Li},  \textsc{F.~Huang},  and  \textsc{L.~Hu},
	\jr{Mater. Sci. Eng. B Solid-State Mater. Adv. Technol.} \textbf{196}(390),
	23--27 (2015).
	
	
	\bibitem{Tian2012}
	\textsc{Y.~Tian},  \textsc{R.~Xu},  \textsc{Y.~Guo},  \textsc{M.~Li},
	\textsc{L.~Hu},  and  \textsc{J.~Zhang},
	\jr{J. Lumin.} \textbf{132}(8), 1873--1878 (2012).
	
	
	\bibitem{Tian2012a}
	\textsc{Y.~Tian},  \textsc{R.~Xu},  \textsc{L.~Hu},  and  \textsc{J.~Zhang},
	\jr{Mater. Lett.} \textbf{69}, 72--75 (2012).
	
	
	\bibitem{Heo1997}
	\textsc{J.~Heo},  \textsc{W.\,Y. Cho},  and  \textsc{W.\,J. Chung},
	\jr{J. Non. Cryst. Solids} \textbf{212}, 151--156 (1997).
	
	
	\bibitem{Antipenko1980a}
	\textsc{B.\,M. Antipenko},  \textsc{A.\,L. Ashkalunin},  \textsc{A.\,A. Mak},
	\textsc{B.\,V. Sinitsyn},  \textsc{Y.\,V. Tomashevich},  and  \textsc{G.\,S.
		Shakhkalamyan},
	\jr{Sov. J. quantum Electron.} \textbf{10}(5), 560--563 (1980).
	
	
	\bibitem{Djeu1997}
	\textsc{N.~Djeu},  \textsc{V.\,E. Hartwell},  \textsc{A.\,A. Kaminskii},  and
	\textsc{A.\,V. Butashin},
	\jr{Opt. Lett.} \textbf{22}(13), 997--999 (1997).
	
	
	\bibitem{Barnes1991}
	\textsc{N.\,P. Barnes} and  \textsc{R.\,E. Allen},
	\jr{IEEE J. Quantum Electron.} \textbf{27}(2), 277--282 (1991).
	
	
	\bibitem{Doroshenko2009}
	\textsc{M.\,E. Doroshenko},  \textsc{T.\,T. Basiev},  \textsc{V.\,V. Osiko},
	\textsc{V.\,V. Badikov},  \textsc{D.\,V. Badikov},
	\textsc{H.~Jel{\'{i}}nkov{\'{a}}},  \textsc{P.~Koranda},  and
	\textsc{J.~Sulc},
	\jr{Opt. Lett.} \textbf{34}(5), 590--2 (2009).
	
	
	\bibitem{Jan2010}
	\textsc{J.~Sulc},  \textsc{H.~Jel{\'{i}}nkov{\'{a}}},  \textsc{M.\,E.
		Doroshenko},  \textsc{T.\,T. Basiev},  \textsc{V.\,V. Osiko},  \textsc{V.\,V.
		Badikov},  and  \textsc{D.\,V. Badikov},
	\jr{Opt. Lett.} \textbf{35}(18), 3051--3053 (2010).
	
	
	\bibitem{Jelinkova2011}
	\textsc{H.~Jelinkova},  \textsc{M.\,E. Doroshenko},
	\textsc{M.~Jel{\'{i}}nek},  \textsc{J.~{\v{S}}ulc},  \textsc{T.\,T. Basiev},
	\textsc{V.\,V. Osiko},  \textsc{V.\,V. Badikov},  and  \textsc{D.\,V.
		Badikov},
	\jr{Laser Phys. Lett.} \textbf{8}(5), 349--353 (2011).
	
	
	\bibitem{Jelinkova2013a}
	\textsc{H.~Jel{\'{i}}nkov{\'{a}}},  \textsc{M.\,E. Doroshenko},
	\textsc{M.~Jel{\'{i}}nek},  \textsc{J.~{\v{S}}ulc},  \textsc{V.\,V. Osiko},
	\textsc{V.\,V. Badikov},  and  \textsc{D.\,V. Badikov},
	\jr{Opt. Lett.} \textbf{38}(16), 3040 (2013).
	
	
	\bibitem{Jackson2003}
	\textsc{S.\,D. Jackson},
	\jr{Appl. Phys. Lett.} \textbf{83}(7), 1316--1318 (2003).
	
	
	\bibitem{Tsang2006a}
	\textsc{Y.\,H. Tsang},  \textsc{A.\,E. El-Taher},  \textsc{T.\,a. King},  and
	\textsc{S.\,D. Jackson},
	\jr{Opt. Express} \textbf{14}(2), 678--685 (2006).
	
	
	\bibitem{Tsang2011}
	\textsc{Y.\,H. Tsang} and  \textsc{A.\,E. El-Taher},
	\jr{Laser Phys. Lett.} \textbf{8}(11), 818--822 (2011).
	
	
	\bibitem{Majewski2016c}
	\textsc{M.\,R. Majewski} and  \textsc{S.\,D. Jackson},
	\jr{Opt. Lett.} \textbf{41}(10), 2173 (2016).
	
	
	\bibitem{Majewski2016}
	\textsc{M.\,R. Majewski} and  \textsc{S.\,D. Jackson},
	\jr{Opt. Lett.} \textbf{41}(19), 4496 (2016).
	
	
	\bibitem{Sojka2018a}
	\textsc{L.~S{\'{o}}jka},  \textsc{L.~Pajewski},  \textsc{M.~Popenda},
	\textsc{E.~Bere{\'{s}}-Pawlik},  \textsc{S.~Lamrini},  \textsc{K.~Markowski},
	\textsc{T.~Osuch},  \textsc{T.\,M. Benson},  \textsc{A.~Seddon},  and
	\textsc{S.~Sujecki},
	\jr{IEEE Photonics Technol. Lett.} \textbf{30}(12), 1083--1086 (2018).
	
	
	\bibitem{Majewski2018a}
	\textsc{M.\,R. Majewski},  \textsc{R.\,I. Woodward},  and  \textsc{S.\,D.
		Jackson},
	\jr{Opt. Lett.} \textbf{43}(5), 971 (2018).
	
	
	\bibitem{Woodward2018b}
	\textsc{R.\,I. Woodward},  \textsc{M.\,R. Majewski},  \textsc{G.~Bharathan},
	\textsc{D.\,D. Hudson},  \textsc{A.~Fuerbach},  \textsc{A.\,S.\,D. Jackson},
	and  \textsc{S.\,D. Jackson},
	\jr{Opt. Lett.} \textbf{43}(7), 1471 (2018).
	
	
	\bibitem{Fortin2019}
	\textsc{V.~Fortin},  \textsc{F.~Jobin},  \textsc{M.~Larose},
	\textsc{M.~Bernier},  and  \textsc{R.~Vall{\'{e}}e},
	\jr{Opt. Lett.} \textbf{44}(3), 491--494 (2019).
	
	
	\bibitem{Majewski2018b}
	\textsc{M.\,R. Majewski},  \textsc{R.\,I. Woodward},  \textsc{J.\,Y. Carree},
	\textsc{S.~Poulain},  \textsc{M.~Poulain},  and  \textsc{S.\,D. Jackson},
	\jr{Opt. Lett.} \textbf{43}(8), 1926 (2018).
	
	
	\othercit
	\bibitem{Sennaroglu2006}
	\textsc{Y.\,H. Tsang},  \textsc{A.\,E. El-Taher},  \textsc{T.\,A. King},
	\textsc{K.\,P. Chang},  and  \textsc{S.\,D. Jackson},
	{Efficient 2.96 micron dysprosium-doped ZBLAN fibre laser pumped at 1.3
		$\mu$m},
	in: Solid State Laser Amplifiers II,  (SPIE, 2006),  pp.\,61900J--61900J--10.
	
	
	\bibitem{Fortin2015}
	\textsc{V.~Fortin},  \textsc{M.~Bernier},  \textsc{S.\,T. Bah},  and
	\textsc{R.~Vall{\'{e}}e},
	\jr{Opt. Lett.} \textbf{40}(12), 2882 (2015).
	
	
	\bibitem{Aydn2018}
	\textsc{Y.\,O. Aydin},  \textsc{V.~Fortin},  \textsc{R.~Vall{\'{e}}e},  and
	\textsc{M.~Bernier},
	\jr{Opt. Lett.} \textbf{43}(18), 4542--4545 (2018).
	
	
	\bibitem{Libatique2000}
	\textsc{N.~Libatique},  \textsc{J.~Tafoya},  \textsc{N.~Viswanathan},
	\textsc{R.\,K. Jain},  and  \textsc{A.~Cable},
	\jr{Electron. Lett.} \textbf{36}(9) (2000).
	
	
	\bibitem{Crawford2015}
	\textsc{S.~Crawford},  \textsc{D.\,D. Hudson},  and  \textsc{S.\,D. Jackson},
	\jr{IEEE Photonics J.} \textbf{7}(3), 1--9 (2015).
	
	
	\bibitem{Woodward2018a}
	\textsc{R.\,I. Woodward},  \textsc{M.\,R. Majewski},  \textsc{D.\,D. Hudson},
	and  \textsc{S.\,D. Jackson},
	\jr{APL Photonics} \textbf{4}, 020801 (2019).
	
	
	\bibitem{Woodward2019}
	\textsc{R.\,I. Woodward},  \textsc{M.\,R. Majewski},  \textsc{N.~Macadam},
	\textsc{G.~Hu},  \textsc{T.~Albrow-Owen},  \textsc{T.~Hasan},  and
	\textsc{S.\,D. Jackson},
	\jr{Opt. Express} \textbf{27}(10), 15032--15045 (2019).
	
	
	\bibitem{Luo2019b}
	\textsc{H.~Luo},  \textsc{J.~Li},  \textsc{Y.~Gao},  \textsc{Y.~Xu},
	\textsc{X.~Li},  and  \textsc{Y.~Liu},
	\jr{Opt. Lett.} \textbf{44}(9), 2322 (2019).
	
	
	\bibitem{Woodward2018}
	\textsc{R.\,I. Woodward},  \textsc{M.\,R. Majewski},  and  \textsc{S.\,D.
		Jackson},
	\jr{APL Photonics} \textbf{3}(11), 116106 (2018).
	
	
	\bibitem{Wang2019}
	\textsc{Y.~Wang},  \textsc{F.~Jobin},  \textsc{S.~Duval},  \textsc{V.~Fortin},
	\textsc{P.~Laporta},  \textsc{M.~Bernier},  \textsc{G.~Galzerano},  and
	\textsc{R.~Vall{\'{e}}e},
	\jr{Opt. Lett.} \textbf{44}(2), 395 (2019).
	
	
	\bibitem{Antipov2016}
	\textsc{S.~Antipov},  \textsc{D.\,D. Hudson},  \textsc{A.~Fuerbach},  and
	\textsc{S.\,D. Jackson},
	\jr{Optica} \textbf{3}(12), 1373 (2016).
	
	
	\bibitem{Woodward2017a}
	\textsc{R.\,I. Woodward},  \textsc{D.~Hudson},  \textsc{A.~Fuerbach},  and
	\textsc{S.~Jackson},
	\jr{Opt. Lett.} \textbf{42}(23), 6--9 (2017).
	
	
	\bibitem{Alamgir2017}
	\textsc{I.~Alamgir},  \textsc{L.~Li},  \textsc{D.\,D. Hudson},
	\textsc{A.~Fuerbach},  \textsc{S.\,D. Jackson},  \textsc{S.~Antipov},
	\textsc{T.~Hu},  \textsc{M.\,E. Amraoui},  \textsc{Y.~Messaddeq},  and
	\textsc{M.~Rochette},
	\jr{Optica} \textbf{4}(10), 1163 (2017).
	
	
	\bibitem{Duval2015}
	\textsc{S.~Duval},  \textsc{M.~Bernier},  \textsc{V.~Fortin},
	\textsc{J.~Genest},  \textsc{M.~Pich{\'{e}}},  and  \textsc{R.~Vall{\'{e}}e},
	\jr{Optica} \textbf{2}(7), 623 (2015).
	
	
	\bibitem{Hu2015}
	\textsc{T.~Hu},  \textsc{S.\,D. Jackson},  and  \textsc{D.\,D. Hudson},
	\jr{Opt. Lett.} \textbf{40}(18), 4226 (2015).
	
	
	\othercit
	\bibitem{Duval2016}
	\textsc{S.~Duval},  \textsc{M.~Olivier},  \textsc{V.~Fortin},
	\textsc{M.~Bernier},  \textsc{M.~Pich{\'{e}}},  and
	\textsc{R.~Vall{\'{e}}e},
	{23-kW peak power femtosecond pulses from a mode-locked fiber ring laser at 2.8
		$\mu$m},
	in: Fiber Lasers XIII Technol. Syst. Appl.,  (SPIE, 2016),  p.\,972802.
	
	
	\bibitem{Zhu2017a}
	\textsc{X.~Zhu},  \textsc{G.~Zhu},  \textsc{C.~Wei},  \textsc{L.\,V. Kotov},
	\textsc{J.~Wang},  \textsc{M.~Tong},  \textsc{R.\,A. Norwood},  and
	\textsc{N.~Peyghambarian},
	\jr{J. Opt. Soc. Am. B} \textbf{34}(3), 15--28 (2017).
	
	
	\bibitem{Sabert1994}
	\textsc{H.~Sabert} and  \textsc{E.~Brinkmeyer},
	\jr{J. Light. Technol.} \textbf{12}(8), 1360--1368 (1994).
	
	
	\bibitem{DeSterke1995}
	\textsc{C.\,M. de~Sterke} and  \textsc{M.\,J. Steel},
	\jr{Opt. Commun.} \textbf{117}(5-6), 469--474 (1995).
	
	
	\bibitem{Majewski2019}
	\textsc{M.\,R. Majewski},  \textsc{R.\,I. Woodward},  and  \textsc{S.\,D.
		Jackson},
	\jr{Opt. Lett.} \textbf{44}(7), 1698 (2019).
	
	
	\othercit
	\bibitem{Quimby2014}
	\textsc{R.\,S. Quimby},
	{Comparison of Fluoroindate and Fluorozirconate Rare Earth Doped Glasses for
		Mid-IR Lasers},
	in: Front. Opt.,  (OSA, 2014).
	
	
	\bibitem{Brunel1996}
	\textsc{M.~Brunel},  \textsc{M.~Vallet},  \textsc{F.~Bretenaker},
	\textsc{A.~Le},  \textsc{J.\,l. Adam},  \textsc{N.~Duhamel-henry},  and
	\textsc{J.\,y. Allain},
	\jr{Opt. Mater. (Amst).} \textbf{5}(March), 209--215 (1996).
	
	
	\othercit
	\bibitem{Quimby2013}
	\textsc{R.\,S. Quimby} and  \textsc{M.~Saad},
	{Dy:fluoroindate Fiber Laser at 4.5 $\mu$m with Cascade Lasing},
	in: Adv. Solid-State Lasers Congr.,  (OSA, 2013),  p.\,AM2A.7.
	
	
	\bibitem{Quimby2008}
	\textsc{R.\,S. Quimby},  \textsc{L.\,B. Shaw},  \textsc{J.\,S. Sanghera},  and
	\textsc{I.\,D. Aggarwal},
	\jr{IEEE Photonics Technol. Lett.} \textbf{20}(2), 123--125 (2008).
	
	
	\bibitem{Sujecki2010}
	\textsc{S.~Sujecki},  \textsc{L.~S{\'{o}}jka},
	\textsc{E.~Bere{\'{s}}-Pawlik},  \textsc{Z.~Tang},  \textsc{D.~Furniss},
	\textsc{A.\,B. Seddon},  and  \textsc{T.\,M. Benson},
	\jr{Opt. Quantum Electron.} \textbf{42}(2), 69--79 (2010).
	
	
	\bibitem{Guo2018}
	\textsc{X.~Xiao},  \textsc{Y.~Xu},  \textsc{H.~Guo},  \textsc{P.~Wang},
	\textsc{X.~Cui},  \textsc{M.~Lu},  \textsc{Y.~Wang},  and  \textsc{B.~Peng},
	\jr{IEEE Photonics J.} \textbf{10}(2), 1--11 (2018).
	
	
	\bibitem{Starecki2015}
	\textsc{F.~Starecki},  \textsc{F.~Charpentier},  \textsc{J.\,L. Doulan},
	\textsc{L.~Quetel},  \textsc{M.~Karine},  \textsc{R.~Chahal},
	\textsc{J.~Troles},  \textsc{B.~Bureau},  \textsc{A.~Braud},  and
	\textsc{P.~Camy},
	\jr{Sens. Actuators, B} \textbf{207}, 518--525 (2015).
	
	
	\bibitem{Ari2018}
	\textsc{J.~Ari},  \textsc{F.~Starecki},  \textsc{C.~Boussard-Pl{\'{e}}del},
	\textsc{Y.~Ledemi},  \textsc{Y.~Messaddeq},  \textsc{J.\,L. Doulan},
	\textsc{A.~Braud},  \textsc{B.~Bureau},  and  \textsc{V.~Nazabal},
	\jr{Opt. Lett.} \textbf{43}(12), 43--46 (2018).
	
	
	\bibitem{Pele2016}
	\textsc{A.\,L. Pel{\'{e}}},  \textsc{A.~Braud},  \textsc{J.\,L. Doualan},
	\textsc{F.~Starecki},  \textsc{V.~Nazabal},  \textsc{R.~Chahal},
	\textsc{C.~Boussard-Pl{\'{e}}del},  \textsc{B.~Bureau},
	\textsc{R.~Moncorg{\'{e}}},  and  \textsc{P.~Camy},
	\jr{Opt. Mater. (Amst).} \textbf{61}, 37--44 (2016).
	
	
	\bibitem{Bernier2007}
	\textsc{M.~Bernier},  \textsc{D.~Faucher},  \textsc{R.~Vall{\'{e}}e},
	\textsc{A.~Saliminia},  \textsc{G.~Androz},  \textsc{Y.~Sheng},  and
	\textsc{S.\,L. Chin},
	\jr{Opt. Lett.} \textbf{32}(5), 454--456 (2007).
	
	
	\bibitem{Bharathan2017}
	\textsc{G.~Bharathan},  \textsc{R.\,I. Woodward},  \textsc{M.~Ams},
	\textsc{D.\,D. Hudson},  \textsc{S.\,D. Jackson},  and  \textsc{A.~Fuerbach},
	\jr{Opt. Express} \textbf{25}(24), 30013 (2017).
	
	
	\othercit
	\bibitem{Stevens2016}
	\textsc{G.~Stevens} and  \textsc{T.~Woodbridge},
	{Mid-IR fused fiber couplers},
	in: Components Packag. Laser Syst. II,  (SPIE, 2016),  p.\,973007.
	
	
	\bibitem{Tavakoli2017}
	\textsc{F.~Tavakoli},  \textsc{A.~Rekik},  and  \textsc{M.~Rochette},
	\jr{IEEE Photonics Technol. Lett.} \textbf{29}(9), 735--738 (2017).
	
	
	\bibitem{Thapa2015}
	\textsc{R.~Thapa},  \textsc{R.\,R. Gattass},  \textsc{V.~Nguyen},
	\textsc{G.~Chin},  \textsc{D.~Gibson},  \textsc{W.~Kim},  \textsc{L.\,B.
		Shaw},  and  \textsc{J.\,S. Sanghera},
	\jr{Opt. Lett.} \textbf{40}(21), 5074 (2015).
	
	
\end{thebibliography}
\end{document}